\documentclass[a4paper,12pt]{iopart}
\usepackage{iopams}
\usepackage{setstack}
\usepackage{graphicx}
\usepackage{bm}
\usepackage{epsfig}
\usepackage{color}

\usepackage{hyperref}

\definecolor{red}{rgb}{0.949,0,0.050}
\newcommand{\diag}{{\rm diag\,}}

\newcommand{\Pf}{{\rm Pf\,}}

\newcommand{\U}{{\rm U}}

\newcommand{\Herm}{{\rm Herm}}

\newcommand{\1}{\leavevmode\hbox{\small1\kern-3.8pt\normalsize1}}
\eqnobysec
\renewcommand{\epsilon}{\varepsilon}
\renewcommand{\bar}[1]{\overline{#1}}

\begin{document}

\title[GUE-chGUE Transition preserving Chirality]{GUE-chGUE Transition  {preserving} Chirality at finite Matrix   {Size}}
\author{Takuya Kanazawa}
\address{Research and Development Group, Hitachi, Ltd., Kokubunji, Tokyo 185-8601, Japan}
\author{Mario Kieburg}
\address{Fakult\"at f\"ur Physik, Universit\"at Bielefeld, 33501 Bielefeld, Germany}
\eads{\mailto{mkieburg@physik.uni-bielefeld.de}}

\begin{abstract}
We study a random matrix model which interpolates between   {the singular values of the} Gaussian unitary ensemble (GUE) and of the  chiral Gaussian unitary ensemble (chGUE). 
This symmetry crossover is analogous to the one realized by the Hermitian Wilson Dirac operator in lattice QCD, but our model preserves chiral symmetry of chGUE exactly unlike the Hermitian Wilson Dirac operator. 
This difference has a crucial impact on the statistics of near-zero eigenvalues, though both   {singular value} statistics build a Pfaffian point process. 
The model in the present work is motivated by the Dirac operator of 3d staggered fermions, 3d QCD at finite isospin chemical potential, and 4d QCD at high temperature. 
We calculate the spectral statistics at finite matrix dimension. For this purpose, we derive the joint probability density of   {the singular values, the} skew-orthogonal polynomials and  {the} kernels for the $k$-point correlation functions. The skew-orthogonal polynomials are constructed by the method of mixing bi-orthogonal and skew-orthogonal polynomials, which is an alternative approach to Mehta's one.  We compare our results with Monte Carlo simulations and study the limits to chGUE and GUE. As a side product, we also  calculate a new type of   {a} unitary group integral.
\end{abstract}
\pacs{02.10.Yn,05.50.+q,11.15.Ex,11.15.Ha,12.38.-t \\MSC numbers: 15B52, 33C45}

\section{Introduction}\label{sec:intro}
  
Random matrix theory (RMT) is a versatile tool for analyzing spectral statistics of operators like Hamiltonians in quantum chaotic and disordered systems~\cite{GMW,Haake,Sieber,Beenakker}, the density operator in quantum information theory~\cite{Bengtsson,ForresterKieburg}, and the Dirac operator in Quantum chromodynamics (QCD)~\cite{Verbaarschot:2000dy,Verbaarschot}. It even allows to compare spectra of completely different systems ranging over many orders of scales. Applications of RMT can also be found beyond physics, like telecommunications, time series analysis in finance, ecology, sociology and medicine, and mathematical topics like algebraic geometry, number theory, combinatorics and graph theory. For more examples, see~\cite{handbook:2010}. In QCD the applicability is two-fold. First, it allows to derive analytical relations between low energy constants of the chiral effective theory (non-linear $\sigma$-models) and spectral observables of the Dirac operator.   {This enables to determine}  {the} low energy constants by lattice simulations; see~\cite{Verbaarschot:2000dy,Verbaarschot}. Second, RMT can be applied to situations where the notorious sign problem impedes lattice simulations, like at finite baryon chemical potential~\cite{Osborn:2004,Splittorff:2006vj,Akemann:2007rf,Kanazawa:2012zzr} or  {at} finite $\theta$-angle~\cite{Damgaard:1999ij,Janik:1999,Kanazawa:2011tt,Kieburg:2017}. 

The random matrix model we consider here is inspired by a certain type of Dirac operators. Hence, it will be of interest in QCD although we   {may expect} applications in other areas, as well. Especially, Hamiltonians in condensed matter theory sometimes share similar or even the same global symmetries as those of Dirac operators in QCD-like theories. Our model is a Gaussian distributed, chiral, two-matrix model exhibiting statistics corresponding to the Dyson index $\beta=2$ in the bulk of the spectrum. The random matrix is explicitly of the form
\begin{equation}\label{RMT-model}
\eqalign{
\fl\mathcal{D} =\Bigg(\begin{array}{cc} 0 & iW \\ iW^\dagger & 0 \end{array}\Bigg),~~~
W= H_1+i\mu H_2,~~~H_1,H_2\in{\rm Herm}(N)\ {\rm and}\ \mu\in[0,1]
 }
\end{equation}
distributed as
\begin{equation}\label{distribution}
P(\mathcal{D})= \frac{1}{2^N \pi^{N^2}}\exp\left[-\frac{1}{2}\Tr(H_1^2+H_2^2)\right]
\end{equation}
with $\Herm(N)$ denoting the set of Hermitian $N\times N$ matrices. The coupling parameter $\mu$ can be chosen real   {in} general. However, due to the symmetries $(\lambda,H_1,H_2,\mu)\to(\lambda,H_1,-H_2,-\mu)$ and $(\lambda,H_1,H_2,\mu)\to(\lambda/\mu,H_2,H_1,1/\mu)$ with $\lambda$ an arbitrary eigenvalue of $\mathcal{D}$, we can reduce its parameter range to $[0,1]$. For $\mu=1$  the model is exactly the one of chGUE while for $\mu=0$ we have the spectral statistics of  {the singular values of the GUE. Hence} the exact chiral pairs of eigenvalues $(\lambda_j,-\lambda_j)$   {are} at any time present.

The model~\eref{RMT-model} is related to the elliptic   {complex Ginibre} ensemble, for which the primary focus has been on the complex eigenvalue spectrum of the matrix $W$~\cite{Sommers,Fyodorov:1997,Akemann:2001bf,Akemann:2007rf}. Its physical application includes the scattering at disordered and chaotic systems~\cite{Fyodorov:1997b}, as well as 3d QCD at finite baryon chemical potential~\cite{Akemann:2001bf,Akemann:2007rf}.   {In comparison to these works}, we are interested in the singular value statistics of $W$. 

There are three applications of \eref{RMT-model} in QCD. The first application is 4d QCD at high temperature. Since the early 80's it is understood that at high temperature QCD-like gauge theories undergo dimensional reduction~\cite{Ginsparg:1980ef,Appelquist:1981vg,Nadkarni:1982kb,Nadkarni:1988fh,Landsman:1989be,Kajantie:1995dw}. In this regime the chiral condensate evaporates and RMT loses its validity for the infrared Dirac spectrum \cite{Kovacs:2009zj}. However, by judiciously choosing the boundary condition of quarks along   {the time-like circle} $S^1$ it is possible to avoid chiral restoration up to an arbitrarily high temperature \cite{Stephanov:1996he,Bilgici:2009tx}. Then the dimensional crossover should manifest itself particularly strong in the smallest eigenvalues of the Dirac operator because they encode the type of spontaneous symmetry breaking.  The Dirac operator of 4d QCD with more than two colors ($N_{\rm c}>2$) and quarks in the fundamental representation shares the global symmetries of chGUE~\cite{Shuryak:1992pi,Verbaarschot:1993pm,Verbaarschot}. In three dimensions the symmetries are those of GUE~\cite{Verbaarschot:1994ip,Akemann:2001bf,Akemann:2001}. Since chiral symmetry has to be always present for the Dirac operator in the 4d continuum theory, we expect the spectral statistics of~\eref{RMT-model}.

The second application can be found in 3d QCD at finite isospin chemical potential $\mu_{\rm I}$ \cite{Akemann:2001bf,Akemann:2007}. When analyzing the pion condensate $\langle\bar{u}d+\bar{d}u\rangle$ ($u$ and $d$ are the up and down quarks), one needs to introduce a source variable $j$. Arranging the two quark fields as $\bar{\psi}=(\bar{u},\bar{d})$, the fermionic part of the Lagrangian reads%
\footnote{The authors of~\cite{Akemann:2001bf,Akemann:2007} did not study the pion condensate $\langle\bar{u}d+\bar{d}u\rangle$. As such, they had no source term $j$ and thus had a decoupling of the two quarks.}
\begin{equation}\label{Lag-phys}
	 \mathcal{L} = \bar{\psi}[\mathcal{D}(m,\mu_{\rm I})+j\tau_1]\psi =\bar{\psi}\bigl[\mathcal{D}_{\rm 3d}
		+ \mu_{\rm I} \sigma_3\tau_3+{\rm diag}(m_{\rm u},m_{\rm d})+j\tau_1\bigl]\psi,
\end{equation}
where $\mathcal{D}_{\rm 3d}$ is the Euclidean anti-Hermitian 3d Dirac operator, $m_{\rm u/d}$ are the quark masses and $\sigma_j$ and $\tau_j$ are the Pauli matrices in spinor and flavor space, respectively.   {Let us} take the chiral limit for simplicity. The resonances (zeros of the characteristic polynomials of $\mathcal{D}+j\tau_1$) in $j$ are the eigenvalues of the operator $-\mathcal{D}(m,\mu_{\rm I})\tau_1$. Nonzero density of the latter at the origin is a necessary condition for the pion condensate formation \cite{Kanazawa:2011tt}. By replacing the operator $\mathcal{D}_{\rm 3d}$ by an anti-Hermitian random matrix  ($i\,\cdot\,$GUE) we arrive at the random matrix model
\begin{equation}\label{Step.model}
	\mathcal{D} = \left(\begin{array}{cc} 0 & iH - {\mu}_{\rm I}\1 \\ 
	iH+ {\mu}_{\rm I}\1 & 0 \end{array}\right),\qquad H=H^\dagger.
\end{equation}
The relation of this model to~\eref{RMT-model} is   {similar to} the relation between the Stephanov model~\cite{Stephanov:1996ki} and the Osborn model~\cite{Osborn:2004} for 4d QCD at finite baryon chemical potential. This means that for large ${\mu}_{\rm I}$ we have a phase transition of the model~\eref{Step.model} to a phase where $\mathcal{D} $   {develops} a spectral gap about the origin. Such a phase does not exist in the model~\eref{RMT-model}. However, in the other phase where the spectral gap is closed, we will show in~\cite{Kanazawa:2018b} that the hard edge statistics at the origin will be the same for both models.

The third application is 3d lattice QCD for staggered fermions. It is known~\cite{Damgaard:1998,Damgaard:2002,Bruckmann:2008,Bialas:2010hb,Kieburg:2014,Kieburg:2017rrk} that symmetries of the staggered lattice Dirac operator do not necessarily agree with those of the continuum theory. Recently a complete classification of such symmetry shift
was given for all dimensions \cite{Kieburg:2017rrk}. Towards the continuum limit the Dirac operator has to undergo a change of symmetries to reach the correct continuum theory. In~\cite{Bialas:2010hb} the model~\eref{RMT-model} was proposed as a description of the symmetry crossover of 3d staggered fermions. The comparison of lattice simulations and Monte Carlo simulations of~\eref{RMT-model} in~\cite{Bialas:2010hb} supports their idea.

Let us mention another model which interpolates between GUE and chGUE, namely of the form
\begin{equation}\label{Wilson}
\eqalign{
\mathcal{D}_5 =\left(\begin{array}{cc} 0 & W \\ W^\dagger & 0 \end{array}\right)+\mu H,
\quad W\in\mathbb{C}^{N\times N}~~{\rm and}~~H\in{\rm Herm}(2N)\,.
 }
\end{equation}
This model was considered in~\cite{Akemann:2011} for the Hermitian Wilson Dirac operator (see also \cite{Damgaard:2010cz,Akemann:2010em} for a related model) {, in particular we want to compare the results of our model with those of the Hermitian Wilson Dirac operator in its chiral limit (only then the spectral gap of $\mathcal{D}_5$ is closed)}. The main difference  {of~\eref{Wilson}} to~\eref{RMT-model} is the loss of chirality. Whenever $\mu\neq0$ there are no exact chiral pairs of eigenvalues like $(\lambda_j,-\lambda_j)$. We will see in sections~\ref{sec:kernel} and~\ref{sec:lim.mu} that this difference has an immediate impact on the behavior of the eigenvalues closest to the origin.

In the present work we will analyze the model~\eref{RMT-model} at finite matrix dimension. For this purpose we first derive the joint probability density of the eigenvalues of $\mathcal{D}$ (equivalent to the singular values of $W$ modulo sign) in section~\ref{sec:jpdf}. To achieve this, we evaluate a unitary group integral of a new kind in~\ref{app:group} which generalizes the Leutwyler-Smilga integral~\cite{Leutwyler:1992yt}. The joint probability density turns out to have a Pfaffian form. This is true also for the model \eref{Wilson} in \cite{Akemann:2011} and several other two-matrix models~\cite{Nagao:1998ysi,Nagao:1999nci,Adler2000,Nagao:2001db,Nagao2007JSP,Forrester2008,Akemann:2009fc,Akemann:2010tv,Kieburg:2012mix,Kieburg-Wilson}, where this list is by no means exhaustive. This Pfaffian form allows us to exploit general results on the method of mixing bi-orthogonal and skew-orthogonal polynomials~\cite{Kieburg:2012mix}. Those results are summarized in~\ref{app:Pfaff} and are used to find the Pfaffian point process (section~\ref{sec:Pfaff}), the kernels (section~\ref{sec:kernel}) and the skew-orthogonal polynomials (section~\ref{sec:poly}). Explicit expressions for the skew-orthogonal polynomials are computed via the supersymmetry method~\cite{Zirnbauer,Guhr}. In section~\ref{sec:lim.mu}, we study the limits to GUE ($\mu=0$) and to chGUE ($\mu=1$) in more detail. The qualitative and quantitative difference between the two models~\eref{RMT-model} and~\eref{Wilson} becomes clearer in the limit $\mu\to0$. We summarize our results in section~\ref{sec:conclusion}.

\section{Joint Probability Distribution}\label{sec:jpdf}

To obtain the eigenvalues of $\mathcal{D}$, see Eq.~\eref{RMT-model},  
we perform the diagonalization $\mathcal{D}=i O(\Lambda\otimes\tau_3)O^\dagger$ with $\Lambda=\diag(\lambda_1,\ldots,\lambda_N)>0$ the singular values of $W$ and $O\in{\rm U}(2N)$. Thus the singular values of $W$ completely determine the eigenvalue spectrum of $\mathcal{D}$. We are interested in the joint probability distribution of $\Lambda$ at finite $N$. For this purpose we express the distribution of $\mathcal{D}$ in terms of $W$ as
\begin{equation}
\fl P(W)=\frac{1}{(2\pi\mu)^{N^2}}\exp\left[-\frac{1+\mu^2}{4\mu^2}\Tr WW^\dagger+\frac{1-\mu^2}{8\mu^2}\Tr(W^2+(W^\dagger)^2)\right].
\end{equation}
To shorten the notation we define
\begin{equation}\label{eq:etadefin}
\eta_{\pm}=\frac{1\pm\mu^2}{4\mu^2}.
\end{equation}
Upon the singular value decomposition $W=U\Lambda V$ the measure transforms as
\cite{Edelman_Rao_2005}
\begin{equation}
\eqalign{
	\label{eq:wmeas}
	 d  W = \frac{2^N \pi^{N^2}}{N!\left(\prod_{j=0}^{N-1}j!\right)^2}
	 d  \mu(U)  d  \mu(V) \Delta_N^2(\Lambda^2) 
	\prod_{i=1}^{N}\lambda_i  d  \lambda_i .
}
\end{equation}
where $d\mu$ is the Haar measure of ${\rm U}(N)$ and 
the differential $dW$ is the product of all independent real differentials of the matrix entries of $W$. 
Hence we have for the joint  probability distribution of $\Lambda$
\begin{equation}
\fl\eqalign{
p(\Lambda)&=\frac{1}{N!\left(\prod_{j=0}^{N-1}2^jj!\right)^2|\mu|^{N^2}}
\Delta_N^2(\Lambda^2)\det\Lambda \exp\left(-\eta_+\Tr \Lambda^2\right)\\
&\quad \times\int_{{\rm U}(N)} d\mu(U)\int_{{\rm U}(N)}d\mu(V)\exp\left[\frac{\eta_-}{2}\Tr ((\Lambda VU)^2+((VU)^\dagger\Lambda)^2)\right]\,.
}
\end{equation}
The integral over $V$ can be absorbed in the integration over $U$. The remaining group integral can be performed with the formula derived in~\ref{app:group}. It is a particular case of the more general group integral
\begin{equation}
	\mathcal{I} \equiv 	\int\limits_{\U(N)} d \mu(U)\; 
	\exp \left[\xi \Tr(AU+U^\dagger B)
	+ \frac{1}{2}\Tr [(AU)^2 + (U^\dagger B)^2] \right],
\end{equation}
with $A$ and $B$ two arbitrary complex matrices and $\xi$ an arbitrary parameter; in our case we have $A=B=\Lambda$ and $\xi=0$. When assuming that the singular values $a=\diag(a_1,\ldots,a_N)$ of $AB$ are non-degenerate, the integral is
\begin{equation}
	\fl\mathcal{I}  = \left(\prod_{j=0}^{N-1}\frac{j!}{\sqrt{4\pi}}\right)
	\frac{e^{-\Tr a^2}}{\Delta_N(a^2)}\left\{\begin{array}{cl} {\Pf}\left[\bm{B}_{\xi}(a_k,a_l)\right]_{k,l=1,\ldots, N}, & N\ {\rm is\ even}, \\ \Pf \left[\begin{array}{c|c} \bm{B}_{\xi}(a_k,a_l) & \bm{C}_\xi(a_k) \\ \hline - \bm{C}_\xi(a_l) & 0 \end{array}\right]_{k,l=1,\ldots,N}, & N\ {\rm is\ odd} \end{array}\right.
\end{equation}
with the functions
\begin{eqnarray}
\eqalign{
	\fl \bm{B}_{\xi}(a_k,a_l) &= 
	\int_{\mathbb{R}^2} d x\,d y~ 
	\frac{x-y}{x+y} 
	\Big[ I_0(2a_k x)I_0(2a_l y)-I_0(2a_l y)I_0(2a_k x) \Big]
	e^{-[(x-\xi)^2+(y-\xi)^2]/2},
	\\
	\fl \bm{C}_\xi(a_k) &= \sqrt{2}\int_{-\infty}^{\infty}d x\; 
	e^{- (x-\xi)^2/2} I_0(2a_k x).
}
\end{eqnarray}
The function $I_0$ is the modified Bessel function of the first kind. These results are also derived in~\ref{app:group}.

The structure of the group integral above carries over to the joint probability density $p(\Lambda)$ which is cast into the following form
\begin{equation}\label{jpdf.Neven}
\eqalign{
p(\Lambda)=&\frac{C_N}{N!}\Delta_N(\Lambda^2)\Pf[G(\lambda_a,\lambda_b)]_{a,b=1,\ldots,N}
}
\end{equation}
for even $N$ and
\begin{equation}\label{jpdf.Nodd}
\eqalign{
p(\Lambda)=&\frac{C_N}{N!}\Delta_N(\Lambda^2)\Pf \!
\left[\begin{array}{c|c} \widetilde{G}(\lambda_a,\lambda_b) & g(\lambda_a) \\ \hline -g(\lambda_b) & 0 \end{array}\right]_{a,b=1,\ldots,N}
}
\end{equation}
for odd $N$. The normalization constant is
\begin{equation}\label{norm.const}
C_N=\prod_{j=0}^{N-1}\frac{1}{\sqrt{4\pi\mu^2}(1-\mu^2)^jj!}
\end{equation}
and the weight functions are
\begin{eqnarray}
\fl	G(\lambda_1,\lambda_2) &=&4\lambda_1\lambda_2e^{-\eta_+(\lambda_1^2+\lambda_2^2)}
		\nonumber\\
\fl		&&\times\int_0^{\pi} d \vartheta\tan \vartheta \sinh\big[ \eta_-(\lambda_1^2-\lambda_2^2)\sin (2\vartheta) \big] 
		\; I_0(2\eta_-\lambda_1\lambda_2\cos (2\vartheta)),
		\label{def.G}\\
\fl	g(\lambda) &=& 2\sqrt{\pi}\lambda e^{-\eta_+\lambda^2} I_0(\eta_-\lambda^2),\label{def.g}\\
\fl	\widetilde{G}(\lambda_1,\lambda_2)&=&G(\lambda_1,\lambda_2)-\frac{g(\lambda_1)}{\bar{g}}H(\lambda_2)+H(\lambda_1)\frac{g(\lambda_2)}{\bar{g}}.\label{def.tildG}
\end{eqnarray}
For the definition of $\widetilde{G}$ we need the integrals
\begin{equation}\label{def.barg}
\bar{g}=\int_0^\infty d\lambda\,\lambda\, g(\lambda) =2\sqrt{\pi}|\mu|=\frac{1}{C_1}
\end{equation}
and
\begin{equation}\label{def.H}
\fl\eqalign{
H(\lambda)&=\int_0^\infty d\lambda' \, \lambda' \, G(\lambda',\lambda)
\\
&=\lambda\; e^{-\eta_+\lambda^2}
		\int_0^{\pi} d \vartheta\tan \vartheta \Biggl(\frac{1}{\eta_+-\eta_-\sin(2\vartheta)}\exp\left[\frac{\eta_--\eta_+\sin(2\vartheta)}{\eta_+-\eta_-\sin(2\vartheta)}\eta_-\lambda^2\right]\\
		&\quad -\frac{1}{\eta_++\eta_-\sin(2\vartheta)}\exp\left[\frac{\eta_-+\eta_+\sin(2\vartheta)}{\eta_++\eta_-\sin(2\vartheta)}\eta_-\lambda^2\right]\Biggl).
}
\end{equation}
The result above resembles the one in~\cite{Akemann:2011} of the Hermitian Wilson Dirac random matrix~\eref{Wilson}.

Let us underline that the joint probability density~\eref{jpdf.Nodd} for odd $N$ can also be written with $\widetilde{G}$ replaced by $G$. Indeed this would be more natural from the perspective of deriving the joint probability density, see~\ref{app:group}. The difference of the two representations is that we subtracted the last row and column from the first $N$ rows and columns which does not change the Pfaffian. In this way the two-point weight $\widetilde{G}$ is orthogonal to the constant, i.e. $\int_0^\infty d\lambda' \,\widetilde{G}(\lambda',\lambda)=0$. The reason why we do this is because we want to pursue the ideas in~\cite{Kieburg:2012mix} regarding the construction of the finite $N$ results via skew-orthogonal polynomials, especially those for odd $N$. This construction differs from Mehta's~\cite[Chapter~5.5.]{Mehta_book} only for odd $N$. It has the advantage that all pairs of skew-orthogonal polynomials can be derived in the same way regardless of the parity of $N$, while in Mehta's construction all polynomials are of the same order since they are modified by the one of the highest order~\cite[Eq.~(5.5.16)]{Mehta_book}, see more in section~\ref{sec:poly} and in~\ref{app:Pfaff}.

\section{Pfaffian Point Process}\label{sec:Pfaff}

The particular Pfaffian form, see Eqs.~\eref{jpdf.Neven} and~\eref{jpdf.Nodd}, of the joint probability density $p(\Lambda)$ implies already quite a lot. For example the partition function $Z_N^{(k_{\rm b},k_{\rm f})}$ 
with $k_{\rm f}$ fermionic quarks and $k_{\rm b}(\leq k_{\rm f})$ bosonic quarks,
\begin{equation}\label{eq:Zval}
\eqalign{
		Z_N^{(k_{\rm b},k_{\rm f})}=& \int d H_1 dH_2 \;P(\mathcal{D})\;\frac{\displaystyle \prod_{j=1}^{k_{\rm f}}\det\Bigg(
			\begin{array}{cc} \kappa_{{\rm f}, j} \1_{N} & iH_1-\mu H_2 \\ iH_1+\mu H_2 & \kappa_{{\rm f}, j}\1_{N} \end{array}\Bigg)}
			{\displaystyle \prod_{j=1}^{k_{\rm b}}\det\Bigg(
			\begin{array}{cc} \kappa_{{\rm b}, j} \1_{N} & iH_1-\mu H_2 \\ iH_1+\mu H_2 & \kappa_{{\rm b}, j}\1_{N} \end{array}\Bigg)}\,,
}
\end{equation}
simplifies drastically. The masses of the bosonic valence quarks must have a non-vanishing real part, ${\rm Re}\,\kappa_{j,{\rm b}}\neq0$, to guarantee the integrability. Usually one sets $k_{\rm b}=k_{\rm f}-N_{\rm f}=k$ and chooses the first $N_{\rm f}$ masses $\kappa_{{\rm f}, j}$ equal to the masses of the dynamical quarks and the remaining $\kappa_{{\rm f}, j}$ and $\kappa_{{\rm b}, j}$ being the valence quark masses which might be complex as it is the case for calculating the $k$-point correlation function.

The partition function~\eref{eq:Zval} can be reduced to a Pfaffian~\cite{Kieburg:2012mix}, see also~\ref{app:Pfaff}. 
When $k_{\rm f}-k_{\rm b}=N_{\rm f}$ is even, we have
\begin{equation}\label{part.finite.even}
\fl\eqalign{
&Z_N^{(k_{\rm b},k_{\rm f})}(\kappa)=\frac{C_N}{C_{N+N_{\rm f}}}\frac{\prod_{a=1}^{k_{\rm b}}\prod_{b=1}^{k_{\rm f}}(\kappa_{{\rm f},b}^2-\kappa_{{\rm b},a}^2)}{\Delta_{k_{\rm b}}(\kappa_{\rm b}^2)\Delta_{k_{\rm f}}(\kappa_{\rm f}^2)}\\
\times&\Pf\left[\begin{array}{c|c} \underset{\ }{\frac{C_{N+N_{\rm f}}}{C_{N+N_{\rm f}+2}}(\kappa_{{\rm b},a}^2-\kappa_{{\rm b},b}^2)Z_{N+N_{\rm f}+2}^{(2,0)}(\kappa_{{\rm b},a},\kappa_{{\rm b},b})} & Z_{N+N_{\rm f}}^{(1,1)}(\kappa_{{\rm b},a},\kappa_{{\rm f},d})/(\kappa_{{\rm f},d}^2-\kappa_{{\rm b},a}^2) \\ \hline -Z_{N+N_{\rm f}}^{(1,1)}(\kappa_{{\rm b},b},\kappa_{{\rm f},c})/(\kappa_{{\rm f},c}^2-\kappa_{{\rm b},b}^2) & \overset{\ }{\frac{C_{N+N_{\rm f}}}{C_{N+N_{\rm f}-2}}(\kappa_{{\rm f},c}^2-\kappa_{{\rm f},d}^2)Z_{N+N_{\rm f}-2}^{(0,2)}(\kappa_{{\rm f},c},\kappa_{{\rm f},d})} \end{array}\right],
}
\end{equation}
where the indices take the values $a,b=1,\ldots,k_{\rm b}$ and $c,d=1,\ldots,k_{\rm f}$. It is worth noting that this representation is valid both for even and odd $N$. The first $N_{\rm f}$ fermionic quark masses can be identified with those of the dynamical quarks, $m_1,\ldots,m_{N_{\rm f}}$. The remaining $k_{\rm b}$ fermionic quark masses, 
as well as those of the bosonic quarks, are from valence quarks. 

To get the corresponding result for odd $N_{\rm f}$, one of the bosonic quark masses has to be taken to infinity yielding a row and a column in the Pfaffian which comprise the partition functions $Z_{N+N_{\rm f}-2}^{(0,1)}(\kappa_{{\rm f},j})$ and $Z_{N+N_{\rm f}+2}^{(1,0)}(\kappa_{{\rm b},j})$, only. The Pfaffian structure~\eref{part.finite.even} carries over to $N\to\infty$ and their expressions in the hard edge limit are given in~\cite{Kanazawa:2018b}.

Another consequence of the Pfaffian form of $p(\Lambda)$ is that the singular values $\Lambda$ build a Pfaffian point process~\cite{Mehta_book}. This means that each $k$-point correlation function,
\begin{equation}
R_{N}^{(k)}(\lambda_1,\ldots,\lambda_k)=\frac{N!}{(N-k)!}\int d\lambda_{k+1}\cdots d\lambda_N \; p(\Lambda),
\end{equation}
can be represented as a $(2k)\times(2k)$ Pfaffian, see~\ref{app:Pfaff},
\begin{equation}
\fl\eqalign{
	R_{N}^{(k)} (\lambda_1,\dots,\lambda_k) =(-1)^{k(k-1)/2} \Pf\left[\begin{array}{c|c}
			W_{N}(\lambda_a,\lambda_b) & G_{N}(\lambda_a,\lambda_c)
			\\ \hline
			- G_{N}(\lambda_d,\lambda_b) & K_{N}(\lambda_d,\lambda_c) \end{array}\right]_{a,b,c,d=1,\ldots,k}.\label{eq:Rkpointev}
}
\end{equation}
The minus sign results from the arrangement of the blocks, namely the upper left corner  only comprises the matrix $W_N$, here. We could also arrange the columns and rows such that each entry consists of a $2\times2$ block containing all three kernels which would absorb the overall sign. Let us underline that the three kernels have a different form for even and odd $N$, as shown in section~\ref{sec:kernel}, while the structure~\eref{eq:Rkpointev} itself does not change.

The normalized level density is given by
\begin{equation}
\rho_N(\lambda)=\frac{1}{N}R_N^{(1)}(\lambda)=\frac{1}{N}G_{N}(\lambda,\lambda).
\end{equation}
We will make use of this relation in section~\ref{sec:kernel}.

The kernels can be given in terms of the three partition functions with two quarks. We have the following formulas~\cite{Kieburg:2012mix}, see also~\ref{app:Pfaff},
\begin{equation}\label{def.kernel}
\fl\eqalign{
K_N(\lambda_1,\lambda_2)=&\frac{C_N}{C_{N-2}}(\lambda_1^2-\lambda_2^2)Z_{N-2}^{(0,2)}(i\lambda_1,i\lambda_2),\\
G_N(\lambda_1,\lambda_2)=&\frac{1}{2\pi}\lim_{\epsilon\to0}\sum_{s=\pm1}s\frac{Z_{N}^{(1,1)}(i\lambda_1+s\epsilon,i\lambda_2)-1}{(\lambda_1-i s\epsilon)^2-\lambda_2^2}, \\
W_N(\lambda_1,\lambda_2)=&\frac{1}{(2\pi)^2}\lim_{\epsilon\to0}\sum_{s_1,s_2=\pm1}s_1s_2\frac{C_N}{C_{N+2}}((\lambda_1-i s_1\epsilon)^2-(\lambda_2-i s_2\epsilon)^2)\\
&\times Z_{N+2}^{(2,0)}(i\lambda_1+ s_1\epsilon,i\lambda_2+ s_2\epsilon).
}
\end{equation}
Again this holds due to the general form of the joint probability density and does not need any detail of the considered model, as can be readily shown by the algebraic rearrangement method proposed in~\cite{KieburgGuhr:2010a,KieburgGuhr:2010b,Kieburg:2012mix}.

We can also include $N_{\rm f}$ dynamical quarks with masses $m_1,\ldots,m_{N_{\rm f}}$ in the $k$-point correlation function~\eref{eq:Rkpointev}. This would yield a shift $N\to N+N_{\rm f}$ in the subscripts of the kernels and, additionally, we would get $N_{\rm f}$ rows and columns comprising $G_{N+N_{\rm f}}(\lambda_a,im_b)$, $K_{N+N_{\rm f}}(\lambda_a,im_b)$ and $K_{N+N_{\rm f}}(im_a,im_b)$. For odd $N_{\rm f}$ we can introduce an additional mass and send it afterwards to infinity. This would give us a further row and column with $\lim_{\epsilon\to0}\sum_{s=\pm1}s Z_{N+N_{\rm f}}^{(1,0)}(i\lambda_a+s\epsilon)/(2\pi)$ and $(C_{N+N_{\rm f}}/C_{N+N_{\rm f}-2})Z_{N+N_{\rm f}}^{(0,1)}(m_a)$.

Concluding this subsection, due to the very particular structure of the joint probability density of the eigenvalues of the Dirac operator $\mathcal{D}$ most quantities can be reduced to the knowledge of only  {a} few functions. How the quantities depend on them is independent of the parity of $N$, only the explicit form of these few functions strongly depends on it.

We would like to point out that similar Pfaffian structures have been derived for several other two matrix models~\cite{Nagao:1998ysi,Nagao:1999nci,Adler2000,Nagao:2001db,Nagao2007JSP,Forrester2008,Akemann:2009fc,Akemann:2010tv,Akemann:2011,Kieburg:2012mix,Kieburg-Wilson}. Considering the fact that determinantal point processes can also be rewritten as Pfaffian ones \cite{Kieburg:2011ct}, Pfaffian point processes seem to be more natural than determinantal point processes.

\section{Skew-Orthogonal Polynomials}\label{sec:poly}

Random matrix ensembles having a probability weight of the form ~\eref{jpdf.Neven} and~\eref{jpdf.Nodd} can generally be solved with the method of 
skew-orthogonal polynomials~\cite{Mehta_book} or a mixed version of bi-orthogonal and skew-orthogonal polynomials~\cite{Kieburg:2012mix} (for odd $N$), see~\ref{app:Pfaff}. 
As we have seen, the \emph{quenched limit} ($N_{\rm f}=0$)~\cite{Verbaarschot:2000dy,Akemann:2007rf} is enough to consider since the theory with dynamical quarks can be easily constructed from it. Therefore we construct only the polynomials corresponding to the quenched weight.

Let us denote by
\begin{equation}
\fl \langle f(WW^\dagger)\rangle_j^{(\alpha,\beta)}=\left(\frac{\alpha^2-\beta^2}{\pi^{2}}\right)^{j^2/2}\int_{\mathbb{C}^{j\times j}}dW\; f(WW^\dagger)\;e^{-\alpha\Tr WW^\dagger+\beta\Tr[W^2+(W^\dagger)^2]/2}
\end{equation}
the average of a function $f$ over a $j\times j$ complex matrix $W$. In this definition the two parameters $\alpha$ and $\beta$ are independent, which is advantageous at a particular step of the calculation below. The random matrix model~\eref{RMT-model} corresponds to $(\alpha,\beta)=(\eta_+,\eta_-)$.

Following the approach in~\cite{Akemann:2010mt}, see also~\ref{app:Pfaff}, we define two kinds of polynomials via Heine-like formulas,
\begin{equation}\label{def:q}
q_j^{(\alpha,\beta)}(x^2)=\langle \det(x^2\1_j-WW^\dagger)\rangle_j^{(\alpha,\beta)}
\end{equation}
and
\begin{equation}\label{def:tildq}
\widetilde{q}_j^{(\alpha,\beta)}(x^2)=\langle 
(x^2+\Tr WW^\dagger+c_j) \det(x^2\1_j-WW^\dagger) \rangle_j^{(\alpha,\beta)}
\end{equation}
with $c_j$ being arbitrary constants which can be adjusted appropriately at the end. The polynomials $q_j^{(\alpha,\beta)}(x^2)$ are of order $j$ in $x^2$ and the polynomials $\widetilde{q}_j^{(\alpha,\beta)}(x^2)$ are of order $j+1$. When we set $(\alpha,\beta)=(\eta_+,\eta_-)$ we choose the short hand notation $q_j=q_j^{(\eta_+,\eta_-)}$ and $\widetilde{q}_j=\widetilde{q}_j^{(\eta_+,\eta_-)}$. Moreover we define the skew-symmetric products
\begin{equation}
\eqalign{
\langle f_1|f_2\rangle_{\rm ev}=&-\langle f_2|f_1\rangle_{\rm ev}=\int_{\mathbb{R}_+^2}d\lambda_1d\lambda_2\; G(\lambda_1,\lambda_2)f_1(\lambda_1)f_2(\lambda_2),\\
\langle f_1|f_2\rangle_{\rm odd}=&-\langle f_2|f_1\rangle_{\rm odd}=\int_{\mathbb{R}_+^2}d\lambda_1d\lambda_2\; \widetilde{G}(\lambda_1,\lambda_2)f_1(\lambda_1)f_2(\lambda_2)\\
}
\end{equation}
for any integrable functions $f_1,f_2$.
The subscripts refer to even and odd $N$.

When using the algebraic rearrangement method in~\cite{KieburgGuhr:2010b}, see also~\ref{app:Pfaff}, we notice that the polynomials are proportional to Pfaffians, cf. Eq.~\eref{pol-construct}. Due to this Pfaffian structure of the polynomials they satisfy the following orthogonality relations by construction (for any $b\in\mathbb{N}_0$)
\begin{equation}
\eqalign{
 \langle \lambda^{2a}|q_{2b}\rangle_{\rm ev}=\langle \lambda^{2a}|\widetilde{q}_{2b}\rangle_{\rm ev}=0,\quad\forall a=0,\ldots,b-1,\\
 \langle \lambda^{2a}|q_{2b+1}\rangle_{\rm odd}=\langle \lambda^{2a}|\widetilde{q}_{2b+1}\rangle_{\rm odd}=0,\quad\forall a=0,\ldots,b,\\
 \int_0^\infty d\lambda\; g(\lambda) q_{2b+1}(\lambda^2)=\int_0^\infty d\lambda \;g(\lambda) \widetilde{q}_{2b+1}(\lambda^2)=0.
}
\end{equation}
This is the foundation of our choice for the skew-orthogonal polynomials in section~\ref{sec:kernel}.

Before proceeding let us find explicit representations for the two kinds of polynomials $q_j^{(\alpha,\beta)}$ and $\widetilde{q}_j^{(\alpha,\beta)}$. We first consider $q_j^{(\alpha,\beta)}$ and follow the ideas of the supersymmetry method~\cite{Zirnbauer,Guhr}. We refer to~\cite{Berezin} for a mathematical introduction to supersymmetry. In the first step we rewrite the determinant as a Gaussian integral over a $j$-dimensional complex Grassmann-valued vector $\psi$,
\begin{equation}
\det(x^2\1_j-WW^\dagger)\propto\int d\psi \; \exp(x^2\psi^\dagger\psi+\Tr WW^\dagger\psi\psi^\dagger).
\end{equation}
We omit the overall constants at the moment since we know that the polynomials are given in monic normalization, $q_j^{(\alpha,\beta)}(x^2)=x^{2j}+\ldots$ In the next step we employ the Hubbard-Stratonovich transformation with a Hermitian matrix $H$
\begin{equation}
\fl e^{\beta\Tr[W^2+(W^\dagger)^2]/2}\propto\int_{{\rm Herm}(j)} d H \exp\left[-\beta\Tr WW^\dagger-\frac{1}{2\beta}\Tr H^2+\Tr H(W+W^\dagger)\right].
\end{equation}
After integration over $W$ we obtain
\begin{equation}
\fl\eqalign{
q_j^{(\alpha,\beta)}(x^2)\propto&\int_{{\rm Herm}(j)} d H\int d\psi\;\det[(\alpha+\beta)\1_j-\psi\psi^\dagger]^{-j}\\
&\times\exp\left[x^2\psi^\dagger\psi-\frac{1}{2\beta}\Tr H^2\left(\1_j-2\beta[(\alpha+\beta)\1_j-\psi\psi^\dagger]^{-1}\right)\right].
}
\end{equation}
The Gaussian integral over $H$ can be performed via the identity
\begin{equation}
\int_{{\rm Herm}(j)} d H \exp(-\Tr H^2 K)\propto\frac{1}{\sqrt{\det(K\otimes\1_j+\1_j\otimes K)}}\;,
\end{equation}
which is valid for any positive definite Hermitian matrix $K$ and can be proven by spectral decomposing $K$ and then integrating $H$ over each matrix entry, separately. It remains to simplify this expression when we set $K=\1_j-2\beta[(\alpha+\beta)\1_j-\psi\psi^\dagger]^{-1}$. For this simplification we make use of the identity
\begin{equation}
\det(A\otimes\1_j+B\otimes\psi\psi^\dagger)=\frac{\det A^{j+1}}{\det(A+B\psi^\dagger\psi)}
\end{equation}
with arbitrary matrices $A$ and $B$, several times. In the end we arrive at
\begin{equation}
\eqalign{
q_j^{(\alpha,\beta)}(x^2)\propto&\int d\psi\;\frac{(\alpha^2-\beta^2-\alpha \psi^\dagger\psi)^{j+1}}{\sqrt{\alpha^2-\beta^2-2\alpha\psi^\dagger\psi+(\psi^\dagger\psi)^2}}\,e^{x^2\psi^\dagger\psi}.
}
\end{equation}
Note that everything depends on $\psi^\dagger\psi$, only. Hence, we can employ the superbosonization formula~\cite{Sommers:2007,Zirnbauer:2008,Kieburg:2008} and replace the integration over $\psi$ by an integration over a phase, $\psi^\dagger\psi\to z$. This yields after proper normalization
\begin{equation}\label{pol.q.gen}
\eqalign{
q_j^{(\alpha,\beta)}(x^2)=&\frac{j!}{(\alpha^2-\beta^2)^{j+1/2}}\oint \frac{d z}{2\pi i\,z^{j+1}}\frac{(\alpha^2-\beta^2-\alpha z)^{j+1}}{\sqrt{\alpha^2-\beta^2-2\alpha z+z^2}}\,e^{x^2z}.
}
\end{equation}
The contour only encircles the origin counter-clockwise. Changing $z\to(\alpha^2-\beta^2) z/\alpha$ we can rewrite the polynomial to 
\begin{equation}
\fl\eqalign{
q_j^{(\alpha,\beta)}(x^2)=&j!\left(\frac{\alpha}{\alpha^2-\beta^2}\right)^j\oint \frac{d z}{2\pi i\,z^{j+1}}\frac{(1- z)^{j}}{\sqrt{1-(\beta/\alpha)^2z^2/(1-z)^2}}\exp\left(\frac{\alpha^2-\beta^2}{\alpha}x^2z\right).
}
\end{equation}
When expanding the square root in $(\beta/\alpha)^2$ we can identify the Laguerre polynomials
\begin{equation}\label{Laguerre}
\eqalign{
	L_k\left(\frac{\alpha^2-\beta^2}{\alpha}x^2\right) &=(-1)^k\oint \frac{d z}{2\pi i\,z^{k+1}}(1- z)^{k}\exp\left(\frac{\alpha^2-\beta^2}{\alpha}x^2z\right)\\
	&=\frac{(-1)^k}{k!}\left(\frac{\alpha^2-\beta^2}{\alpha}x^2\right)^k+\ldots
}
\end{equation}
This yields a more explicit expression in terms of a finite sum,
\begin{equation}
\eqalign{
		q_{j}^{(\alpha,\beta)}(x^2) = & j! \left(\frac{\alpha}{\alpha^2-\beta^2}\right)^j
		\sum_{l=0}^{\lfloor j/2 \rfloor}
		\Bigg(\begin{array}{c} 2l \\ l \end{array}\Bigg)
		\Big(\frac{\beta}{2\alpha}\Big)^{2l}L_{j-2l}
		\left(\frac{\alpha^2-\beta^2}{\alpha}x^2\right).
		\label{eq:L3r}
}
\end{equation}
 {Here, we have used the floor function $\lfloor j/2 \rfloor$ yielding the largest integer which is smaller than or equal to $j/2$.}

An expression analogous to~\eref{pol.q.gen} can be derived for $\widetilde{q}_j^{(\alpha,\beta)}$. In fact it can be completely derived from $q_j^{(\alpha,\beta)}$. Recalling the definition~\eref{def:tildq} we notice that the term $x^2+c_j$ can be pulled out of the integral such that these terms are proportional to $q_j^{(\alpha,\beta)}$. The term with $\Tr WW^\dagger$ can be generated by a derivative in $\alpha$. However we have, then, also to differentiate the normalization constant but this yields only a shift in the arbitrary constants $c_j$. With a slight abuse of notation, we denote the new constants also by $c_j$. We have
\begin{equation}
\widetilde{q}_j^{(\alpha,\beta)}(x^2)=\left(-\partial_\alpha+x^2+c_j\right)q_j^{(\alpha,\beta)}(x^2).
\end{equation}
When applying this relation to the result~\eref{pol.q.gen}, we find (after  {shifting $c_j$ again})
\begin{equation}\label{pol.tildq.gen}
\fl\eqalign{
\widetilde{q}_j^{(\alpha,\beta)}(x^2)=&\frac{j!}{(\alpha^2-\beta^2)^{j+1/2}}\oint \frac{d z}{2\pi i\,z^{j+1}}\frac{(\alpha^2-\beta^2-\alpha z)^{j}}{(\alpha^2-\beta^2-2\alpha z+z^2)^{3/2}}\,e^{x^2z}\\
&\times\bigl[(\alpha^2-\beta^2-\alpha z)(\alpha-z)+(\alpha^2-\beta^2-2\alpha z+z^2)\\
&\times(-(j+1)(2\alpha-z)+(x^2+c_j)(\alpha^2-\beta^2-\alpha z))\bigl].
}
\end{equation}
In terms of the Laguerre polynomials this expression reads
\begin{equation}
\fl\eqalign{
&\widetilde{q}_j^{(\alpha,\beta)}(x^2)=\tilde{c}_j q_j^{(\alpha,\beta)}(x^2)+ j! \left(\frac{\alpha}{\alpha^2-\beta^2}\right)^{j+1}
		\sum_{l=0}^{\lfloor j/2 \rfloor}
		\Bigg(\begin{array}{c} 2l \\ l \end{array}\Bigg)
		\Big(\frac{\beta}{2\alpha}\Big)^{2l} 
		\\
		&\qquad \times\Bigg[
			\frac{\beta^2}{\alpha^2}(j-2l) L_{j-2l-1}\left(\frac{\alpha^2-\beta^2}{\alpha}x^2\right)  - (j-2l+1) 
			L_{j-2l+1}\left(\frac{\alpha^2-\beta^2}{\alpha}x^2\right)
		\Bigg]
		\label{eq:L3r2}
}
\end{equation}
after an additional shift of the constant from $c_j$ to $\tilde{c}_j$.
Here we used the identities $\partial L_n(x)=n[L_n(x)-L_{n-1}(x)]/x$ and $xL_n(x)=(2n+1)L_n(x)-nL_{n-1}(x)-(n+1)L_{n+1}(x)$.

Both results, Eqs.~\eref{pol.q.gen} and~\eref{pol.tildq.gen}, simplify when setting $(\alpha,\beta)=(\eta_+,\eta_-)$.  {Thus,} we arrive at the main results of this subsection,
\begin{equation}\label{pol.q}
q_j(x^2)=j!\oint \frac{d z}{2\pi i\,z^{j+1}}\frac{(1-(1+\mu^2)z)^{j+1}}{\sqrt{1-2(1+\mu^2) z+4\mu^2z^2}}e^{x^2z}
\end{equation}
and
\begin{equation}\label{pol.tildq}
\fl\eqalign{
\widetilde{q}_j(x^2)=\;&j!\oint \frac{d z}{2\pi i\,z^{j+1}}\frac{(1-(1+\mu^2)z)^{j}}{(1-2(1+\mu^2) z+4\mu^2z^2)^{3/2}}e^{x^2z}\\
&\times\bigl[(1-(1+\mu^2)z)(1+\mu^2-4\mu^2z)+(1-2(1+\mu^2) z+4\mu^2z^2)\\
&\times(-(j+1)(2(1+\mu^2)-4\mu^2 z)+(x^2+c_j)(1-(1+\mu^2)z))\bigl].
}
\end{equation}
When we define the quotient
\begin{equation}
h_j=\frac{C_{j}}{C_{j+2}}=4\pi\mu^2(1-\mu^2)^{2j+1}j!(j+1)!
\end{equation}
with $C_{-n}=1$ for $n\in\mathbb{N}_0$, each pair $(q_j,\widetilde{q}_j)$ satisfies the normalization
\begin{equation}
\fl\eqalign{
\langle q_{2j}|\widetilde{q}_{2j}\rangle_{\rm ev}=\langle \lambda^{4j}|\widetilde{q}_{2j}\rangle_{\rm ev}=h_{2j}\quad{\rm and}\quad\langle q_{2j+1}|\widetilde{q}_{2j+1}\rangle_{\rm odd}=\langle \lambda^{4j+2}|\widetilde{q}_{2j+1}\rangle_{\rm odd}=h_{2j+1}.
}
\end{equation}
This can be readily checked by the Pfaffian representation~\eref{pol-construct} of the polynomials.

Now we are well-prepared for giving explicit representations of the kernels~\eref{def.kernel} since the partition functions for two flavors are directly given in terms of the skew-orthogonal polynomials, see~\cite{Kieburg:2012mix} and~\ref{app:Pfaff}.

\section{Kernels}\label{sec:kernel}

The skew-orthogonal polynomials are different for even and odd $N$ because the two-point weight changes, cf. Eqs.~\eref{def.G} and~\eref{def.tildG}. Therefore also the explicit form of the kernels~\eref{def.kernel} will be different. We collect the results for even $N$ in subsection~\ref{sec:kernel.even} and for odd $N$ in subsection~\ref{sec:kernel.odd}.

\subsection{Even $N$}\label{sec:kernel.even}

\begin{figure}[t]
	\centering 
	\includegraphics[width=\textwidth]{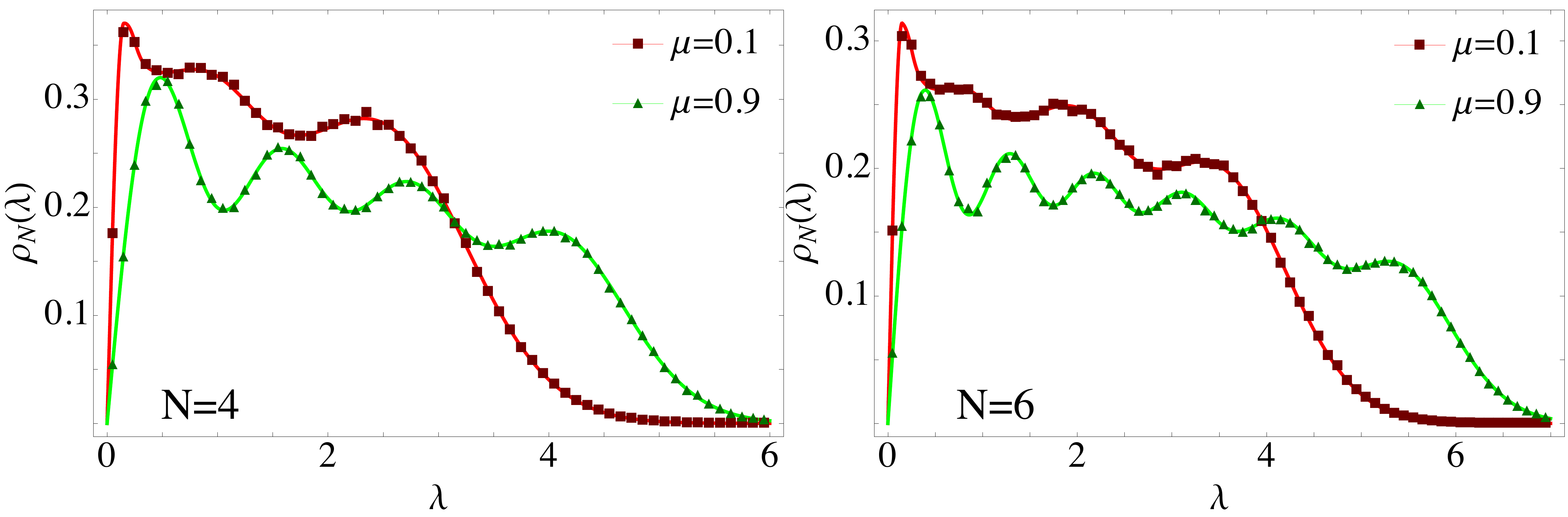}
	\caption{\label{fig:even} The normalized level density
	 for the quenched ensemble  of even, small matrix size ($N=4$, left plot, and $N=6$, right plot).  The analytical result~\eref{level.density.even} (solid curves) are compared with Monte Carlo simulations (symbols). We plotted the ensemble for the coupling parameter $\mu=0.1$ (red squares) and $\mu=0.9$ (green triangles). The ensemble of the Monte Carlo simulations consists of $10^5$ matrices drawn from the random matrix model~\eref{RMT-model}.}
\end{figure}

For even $N$ the skew-orthogonal polynomials and their normalization constants are given by the triple $\{q_{2j},\widetilde{q}_{2j},h_{2j}\}_{j=0,\ldots,N/2-1}$, see~\ref{app:Pfaff}. Thus the kernels for the $k$-point correlation function~\eref{eq:Rkpointev} are given by
\begin{equation}\label{kernel.even}
\eqalign{
W_{N}(\lambda_1,\lambda_2)=&-G(\lambda_1,\lambda_2)-\int_{\mathbb{R}_+^2}dx_1dx_2\;G(x_1,\lambda_1)G(x_2,\lambda_2)K_{N}(x_1,x_2),\\
G_{N}(\lambda_1,\lambda_2)=&\int_0^\infty dx\; G(x,\lambda_1)K_{N}(x,\lambda_2),\\
K_{N}(\lambda_1,\lambda_2)=&\sum_{j=0}^{N/2-1}\frac{q_{2j}(\lambda_2^2)\widetilde{q}_{2j}(\lambda_1^2)-q_{2j}(\lambda_1^2)\widetilde{q}_{2j}(\lambda_2^2)}{4\pi\mu^2(1-\mu^2)^{4j+1}(2j)!(2j+1)!}.
}
\end{equation}
The level density at finite $N$ is then
\begin{equation}\label{level.density.even}
\rho_N(\lambda)=\int_0^\infty dx\;G(x,\lambda)\sum_{j=0}^{N/2-1}\frac{q_{2j}(\lambda^2)\widetilde{q}_{2j}(x^2)- q_{2j}(x^2)\widetilde{q}_{2j}(\lambda^2)}{4\pi\mu^2(1-\mu^2)^{4j+1}(2j)!(2j+1)!}.
\end{equation}
We show its behavior in Fig.~\ref{fig:even} and compare it with Monte Carlo simulations of the model~\eref{RMT-model} for small $N$.

It is notable that for decreasing $\mu$ a discontinuity of the level density is building up  at the origin. The reason for this is that the limit $\mu\to0$ is not uniform, see also~\cite{Bialas:2010hb} where it was observed for the level density of the staggered Dirac operator in three dimensions. This can be understood by the level densities of  {the} GUE and  {the} chGUE. While the level density of  {the} GUE is non-zero at the origin it vanishes linearly for the chGUE, see~\cite{Mehta_book}.

Another important point is the approach to the limit $\mu\to0$ compared to the GUE and  {the} chGUE interpolation in~\cite{Akemann:2011,Damgaard:2010cz,Akemann:2010em} where chirality is broken. In~\cite{Akemann:2011} the authors considered the random matrix model~\eref{Wilson}. The particular form of this model implies that regardless of how small $\mu$ is the chirality is broken and one has a finite density at the origin. In our model~\eref{RMT-model} we preserve chirality which implies that we have always a linear drop off at the origin. The level repulsion reflected in this behavior results from the exact chiral pairs $(\lambda_j,-\lambda_j)$ of eigenvalues of $\mathcal{D}$ which feel each other and which is missing in the model~\eref{Wilson}. The regime were the interaction of the chiral pairs $(\lambda_j,-\lambda_j)$ takes place is of order $\mu$ for small $\mu$ and shows up in the level density about the origin, see Fig.~\ref{fig:even}.

The third point we want to emphasize is the merging of eigenvalue peaks of $\mathcal{D}$ for $\mu\to0$ on the positive and negative line, cf. Fig.~\ref{fig:even}. The reason is that we have on average only $N/2$ eigenvalues on the positive and negative axis, separately, for GUE. Those are represented by $N/2$ peaks in the level density. For chGUE  we have $N$ peaks, thus, twice as much. This is also the reason why the width of the level density for $\mu\approx1$ is obviously bigger than the one for $\mu\approx0$ {; we note that the level density is always normalized to unity.} One can interpret this behavior also differently. Since we plot in Fig.~\ref{fig:even} the singular values of $\mathcal{D}$ one has to compare it with the level density of the eigenvalues of GUE while the singular values of the GUE is equivalent to a direct sum of two independent random matrices, see~\cite{Akemann:2001,Forrester:2006,Edelman:2014,Bornemann:2016}.

In comparison to the limit $\mu\to0$, the limit $\mu\to1$ seems to be less dramatic. The level density approaches this limit uniformly without any surprising features.

\subsection{Odd $N$}\label{sec:kernel.odd}

\begin{figure}[t]
	\centering 
	\includegraphics[width=\textwidth]{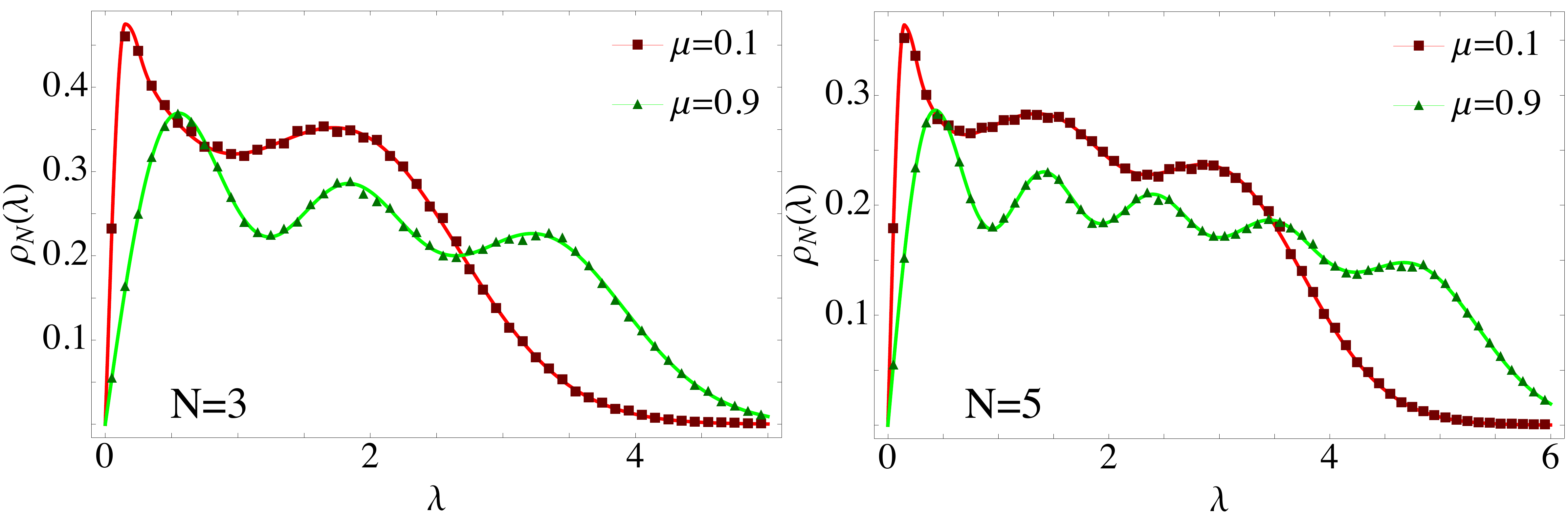}
	\caption{\label{fig:odd} The normalized level density
	 for the quenched ensemble at odd matrix dimension $N=3$ (left plot) and $N=5$ (right plot).  Again we compare the analytical result~\eref{level.density.odd} (solid curves) with Monte Carlo simulations (symbols). The coupling parameter is, as before, $\mu=0.1$ (red squares) and $\mu=0.9$ (green triangles). For the Monte Carlo simulations we have drawn $10^5$ matrices  from the random matrix model~\eref{RMT-model} such that the statistical error is about one percent.}
\end{figure}

Let us consider the case of odd $N$, now. Then, the skew-orthogonal polynomials and their normalizations are $\{q_{2j+1},\widetilde{q}_{2j+1},h_{2j+1}\}_{j=0,\ldots,(N-3)/2}$, see~\ref{app:Pfaff}. We underline that the polynomial of order zero, which is $1$, is not missing. It corresponds to the one-point weight $g(\lambda)$, see Eq.~\eref{def.g}. Therefore the kernels have the form
\begin{equation}\label{kernel.odd}
\fl\eqalign{
W_{N}(\lambda_1,\lambda_2)=&-\widetilde{G}(\lambda_1,\lambda_2)-\int_{\mathbb{R}_+^2}dx_1dx_2\; \widetilde{G}(x_1,\lambda_1)\widetilde{G}(x_2,\lambda_2)K_{N}(x_1,x_2),\\
G_{N}(\lambda_1,\lambda_2)=&\frac{\lambda_1 }{|\mu|}e^{-\eta_+\lambda_1^2} I_0(\eta_-\lambda_1^2)+\int_0^\infty dx \; \widetilde{G}(x,\lambda_1)K_{N}(x,\lambda_2),\\
K_{N}(\lambda_1,\lambda_2)=&\sum_{j=0}^{(N-3)/2}\frac{q_{2j+1}(\lambda_2^2)\widetilde{q}_{2j+1}(\lambda_1^2)-q_{2j+1}(\lambda_1^2)\widetilde{q}_{2j+1}(\lambda_2^2)}{4\pi\mu^2(1-\mu^2)^{4j+3}(2j+1)!(2j+2)!}.
}
\end{equation}
We want to point out the additional term in $G_N$ in comparison to~\eref{kernel.even} which results from $g(\lambda)$ and essentially describes the eigenvalue closest to the origin.
The formulas~\eref{kernel.odd} imply the level density
\begin{equation}\label{level.density.odd}
\fl\eqalign{
\rho_N(\lambda)=\;&\frac{\lambda }{|\mu|}e^{-\eta_+\lambda^2} I_0(\eta_-\lambda^2)\\
&+\int_0^\infty dx \; \widetilde{G}(x,\lambda)\sum_{j=0}^{(N-3)/2}\frac{q_{2j+1}(\lambda^2)\widetilde{q}_{2j+1}(x^2)-  {q_{2j+1}(x^2)}\widetilde{q}_{2j+1}(\lambda^2)}{4\pi\mu^2(1-\mu^2)^{4j+3}(2j+1)!(2j+2)!}.
}
\end{equation}
Its behavior and the comparison with Monte Carlo simulations are displayed in Fig.~\ref{fig:odd}.

The behavior of the limits $\mu\to0,1$ of the level density~\eref{level.density.odd} is more or less the same as for even $N$. The only difference is the number of peaks. While for even $N$ the density converges to a distribution with $N/2$ peaks in the limit $\mu\to0$, the number is $(N+1)/2$ for odd $N$. At the origin an unpaired peak merges with the one on the negative axis. This merging is non-uniform as we can see in Fig.~\ref{fig:odd}. The reason is the same as for the even case and has its origin in the preserved chirality, which is completely different from the results in~\cite{Akemann:2011}, cf. Eq.~\eref{Wilson}, where chirality is broken.

\section{\boldmath Limits $\mu\to0$ and $\mu\to1$ at finite $N$}\label{sec:lim.mu}

As we have already seen in section~\ref{sec:kernel}, the limits $\mu\to0$ and $\mu\to1$ are differently approached. In this section we want to analytically understand how they are approached. For this purpose we consider two quantities. The first are the polynomials $q_j$, see~\eref{def:q}, and the second is the joint probability density of the singular values $\Lambda$, see Eqs.~\eref{jpdf.Neven} and~\eref{jpdf.Nodd}. The limit $\mu\to0$ is analyzed in subsection~\ref{sec:lim.mu0} and the limit $\mu \to 1$ in subsection~\ref{sec:lim.mu1}.

\subsection{Limit $\mu\to0$}\label{sec:lim.mu0}

We want to consider the limit $\mu\to0$ for the polynomial $q_j$, see Eq.~\eref{pol.q}, which is the average of a characteristic polynomial. When setting $\mu=0$ we have the average
\begin{equation}
q_j(x^2,\mu=0)=\langle\det(x^2\1_j-H^2)\rangle_{\rm GUE},
\end{equation}
which is an average over a $j\times j$ dimensional GUE matrix $H$.
Since  { the GUE yields} a determinantal point process we already know the answer~\cite{Mehta_book}:
\begin{equation}\label{pol.q.mu0}
\eqalign{
q_j(x^2,\mu=0)&=(-1)^j\frac{H_{j+1}(x)H_{j}(-x)-H_{j+1}(-x)H_{j}(x)}{2x}\\
&=(-2)^j\left\lceil\frac{j}{2}\right\rceil\left\lfloor\frac{j}{2}\right\rfloor L_{\lceil j/2\rceil}^{(-1/2)}\left(\frac{x^2}{2}\right)L_{\lfloor j/2\rfloor}^{(+1/2)}\left(\frac{x^2}{2}\right)
}
\end{equation}
with $H_j$ the monic Hermite polynomials with respect to the weight $e^{-x^2/2}$, $L_j^{(\nu)}$ being the generalized Laguerre polynomials and we have employed the floor ($\lfloor.\rfloor$) and  {the} ceil ($\lceil.\rceil$) function. This result can be checked with the help of Eq.~\eref{eq:L3r} with $(\alpha,\beta)=(\eta_+,\eta_-)$ at $\mu=0$.

From the result~\eref{pol.q.mu0} one can guess that the singular value statistics  of the ensemble is factorizing for $\mu\to0$ as it is indeed known for the singular values of GUE, see~\cite{Akemann:2001,Forrester:2006,Edelman:2014,Bornemann:2016}. The joint probability density of the eigenvalues $E$   {of the GUE} is given by~\cite{Mehta_book}
\begin{equation}\label{jpdf.ev.GUE}
p_{\rm GUE}^{\rm (ev)}(E)=\frac{1}{N!}\left(\prod_{j=0}^{N-1}\frac{1}{\sqrt{2\pi}j!}\right)\Delta_N^2(E)\exp\left(-\frac{1}{2}\Tr E^2\right).
\end{equation}
The singular values $\Lambda$ are the modulus of the eigenvalues, i.e. $\lambda_j=|E_j|$. Hence we have to sum over the signs of the eigenvalues. Since the Gaussian is even, the monomials of the two Vandermonde determinants can only combine as even order with even order and odd with odd. Therefore the joint probability density of the singular values $\Lambda$ of a matrix $H$ drawn from a GUE is~\cite{Akemann:2001,Forrester:2006,Edelman:2014,Bornemann:2016}
\begin{equation}\label{jpdf.sv.GUE}
\fl p_{\rm GUE}^{\rm (sv)}(\Lambda)=\left(\prod_{j=0}^{N-1}\frac{2}{\sqrt{2\pi}j!}\right)\frac{\Delta_{\lceil N/2\rceil}^2(\Lambda_{\rm ev}^2)}{(\lceil N/2\rceil)!}\frac{\Delta_{\lfloor N/2\rfloor}^2(\Lambda_{\rm odd}^2)}{(\lfloor N/2\rfloor)!}\det\Lambda_{\rm odd}^2\exp\left(-\frac{1}{2}\Tr \Lambda^2\right),
\end{equation}
where $\Lambda=\diag(\Lambda_{\rm ev},\Lambda_{\rm odd})$, $\Lambda_{\rm ev}=\diag(\lambda_1,\ldots,\lambda_{\lceil N/2\rceil})$, and $\Lambda_{\rm odd}=\diag(\lambda_{\lceil N/2\rceil+1},\ldots,\lambda_{N})$. Hence it is a sum of two complex Laguerre ensembles, one with index $-1/2$ and of dimension $\lceil N/2\rceil$ and the other one with $+1/2$ and dimension $\lfloor N/2\rfloor$. These two Laguerre ensemble correspond to the Gaussian antisymmetric unitary ensemble of even and odd dimension (GAOE, see~\cite{Kieburg:2017rrk} for the notation), meaning Gaussian distributed imaginary anti-symmetric matrices. This factorization was also observed in section~\ref{sec:kernel}.

From this picture it becomes clear what the level density of $\Lambda$ is. It is the sum of the level densities of $\Lambda_{\rm ev}$ and $\Lambda_{\rm odd}$. This is in agreement with the level density of GUE, see~\cite{Mehta_book,Akemann:2001,Forrester:2006,Edelman:2014,Bornemann:2016}, because Hermite polynomials of even order, $H_{2j}$, can be expressed in terms of the generalized Laguerre polynomials $L_{j}^{(-1/2)}$ and those of odd order, $H_{2j+1}$, by the Laguerre polynomials $L_{j}^{(+1/2)}$.

We can derive the result above from the joint probability distributions~\eref{jpdf.Neven} and~\eref{jpdf.Nodd} by considering the asymptotics of the two-point weight
\begin{equation}
G(\lambda_1,\lambda_2)\overset{|\mu|\ll1}{\propto}\sum_{s_1,s_2=\pm1}\frac{(s_1\lambda_1-s_2\lambda_2)^2}{\lambda_1^2-\lambda_2^2}\exp\left(-\frac{\lambda_1^2+\lambda_2^2}{2}\right)
\end{equation}
and the one-point weight
\begin{equation}
g(\lambda)\overset{|\mu|\ll1}{\propto}\exp\left(-\frac{\lambda^2}{2}\right).
\end{equation}
The asymptotics of the two-point weight can be found by noticing that the saddle points of the integral in Eq.~\eref{def.G} are given by $(\cos[\vartheta+\pi/4],\sin[\vartheta+\pi/4])=(\pm\lambda_2,\pm\lambda_1)/\sqrt{\lambda_1^2+\lambda_2^2}$ for one term of the {\em sine hyperbolic} and $(\cos[\vartheta+\pi/4],\sin[\vartheta+\pi/4])=(\pm\lambda_1,\pm\lambda_2)/\sqrt{\lambda_1^2+\lambda_2^2}$ for the other term. The Gaussian terms can be pulled out of the Pfaffian and for the remaining term we use
\begin{equation}
\fl\Pf\left[\sum_{s_a,s_b=\pm1}\frac{(s_a\lambda_a-s_b\lambda_b)^2}{\lambda_a^2-\lambda_b^2}\right]_{a,b=1,\ldots,N}=\pm \sum_{s_1,\ldots,s_N=\pm1}\frac{\Delta_N^2(s_1\lambda_1,\ldots,s_N\lambda_N)}{\Delta_N(\Lambda^2)}
\end{equation}
for even $N$ and similar for odd $N$. The Vandermonde determinant in the denominator cancels with those in Eqs.~\eref{jpdf.Neven} and~\eref{jpdf.Nodd}. The sum over the signs and the regrouping of the singular values into $\Lambda_{\rm ev}$ and $\Lambda_{\rm odd}$ yield the expected result~\eref{jpdf.sv.GUE}.

This kind of limit is in the sense of~\cite{ForresterKieburg} where the Pfaffian structure results from the Schur Pfaffian identity~\cite{Schur}
\begin{equation}\label{Pfaff-ident}
\frac{\Delta_{N}^2(\Lambda)}{\Delta_{N}(\Lambda^2)}=\left\{\begin{array}{cl} \displaystyle \Pf\left[\frac{\lambda_b-\lambda_a}{\lambda_b+\lambda_a}\right]_{a,b=1,\ldots,N}, & {\rm for}\ N\ {\rm even}, \\ \Pf\left[\begin{array}{c|c} 0 & 1\ \cdots\ 1  \\ \hline \begin{array}{c} -1 \\ \vdots \\ -1 \end{array} & \displaystyle\frac{\lambda_b-\lambda_a}{\lambda_b+\lambda_a} \end{array}\right]_{a,b=1,\ldots,N}, &  {\rm for}\ N\ {\rm odd}. \end{array}\right.
\end{equation}

\subsection{Limit $\mu\to1$}\label{sec:lim.mu1}

As before we first consider the polynomial $q_j$, see Eq.~\eref{pol.q}, because it is simpler in interpretation. The limit $\mu\to1$ corresponds to chGUE. Thus the averaged characteristic polynomial is proportional to the Laguerre polynomial $L_{j}(x^2/2)$ due to the scaling of the distribution~\eref{distribution}. Indeed when setting $\mu=1$ in Eq.~\eref{pol.q} we have
\begin{equation}\label{pol.q.mu1}
q_j(x^2,\mu=0)=j!\oint \frac{d z}{2\pi i\,z^{j+1}}(1-2z)^{j}e^{x^2z}=(-2)^j j!L_{j}\left(\frac{x^2}{2}\right)
\end{equation}
confirming our expectations. We used Eq.~\eref{Laguerre} in the second equality.

We can derive the limit $\mu\to1$ to chGUE also on the level of the joint probability densities~\eref{jpdf.Neven} and~\eref{jpdf.Nodd}. This time we expand the two-point weight as
\begin{equation}
G(\lambda_1,\lambda_2)\overset{|\mu|\approx1}{\propto}\exp\left(-\frac{\lambda_1^2+\lambda_2^2}{2}\right)\sum_{a,b=0}^\infty G_{ab}(1-\mu^2)^{a+b}\lambda_1^{2a+1}\lambda_2^{2b+1}
\end{equation}
with $G_{ab}$ antisymmetric while the one-point weight is
\begin{equation}
g(\lambda)\overset{|\mu|=1}{\propto}\lambda\exp\left(-\frac{\lambda^2}{2}\right).
\end{equation}
Due to the skew-symmetry of the Pfaffian one can start the series of $G(\lambda_1,\lambda_2)$ with $a$ and $b$ at $1$ when $N$ is odd. The skew-symmetry is also the reason why we cannot just take the leading order term in $(1-\mu^2)$ but need the series which can be minimally cut off at $a,b=N-1$. Pulling the factors $\lambda_j\exp\left[-\lambda_j^2/2\right] $ out the Pfaffian and defining
\begin{equation}
\mathcal{G}=\{G_{ab}\}\quad{\rm and}\quad \mathcal{V}=\{\lambda_b^{2a}\},
\end{equation}
the double sum is equal to $\mathcal{V}^T\mathcal{G}\mathcal{V}$. To evaluate the Pfaffian we can employ
\begin{equation}\label{Pfaff-Vand}
\Pf[\mathcal{V}^T\mathcal{G}\mathcal{V}]=\det \mathcal{V}\ \Pf \mathcal{G}
\end{equation}
for even $N$ (and similar for odd $N$). The Pfaffian $\Pf \mathcal{G}$ is a constant while the determinant of the Vandermonde matrix $\mathcal{V}$ is equal to the Vandermonde determinant $\det\mathcal{V}=\Delta_N(\Lambda^2)$. This term together with the other Vandermonde determinant in Eqs.~\eref{jpdf.Neven} and~\eref{jpdf.Nodd} and the product $\prod_{j=1}^N\lambda_j\exp\left[-\lambda_j^2/2\right]$ yields the joint probability density of the eigenvalues of the Laguerre ensemble of index $0$, see~\cite{Mehta_book}
\begin{equation}\label{jpdf.ev.LUE}
p_{\rm Lag}(\Lambda)=\frac{1}{N!}\left(\prod_{j=0}^{N-1}\frac{1}{2^{2j}(j!)^2}\right)\Delta_N^2(\Lambda^2)\det\Lambda\exp\left(-\frac{1}{2}\Tr \Lambda^2\right).
\end{equation}
This limit arises in a way as proposed in~\cite{Kieburg:2011ct} where a non-trivial Pfaffian is created by rephrasing the Vandermonde determinant as in~\eref{Pfaff-Vand}, which has to be seen in comparison to subsection~\ref{sec:lim.mu0} for $\mu\to0$ where the Pfaffian arises in a totally different way from a determinantal point process.

\section{Conclusions}\label{sec:conclusion}

We computed the joint probability density of the chiral random matrix model~\eref{RMT-model} and studied its eigenvalue statistics at finite matrix dimension $N$. The statistics are governed by a Pfaffian point process meaning that all observables depending on the eigenvalues of the chiral random matrix $\mathcal{D}$, only, can be expressed in terms of a small number of functions, the kernels. This also carries over to the limit of large matrix dimension, see~\cite{Kanazawa:2018,Kanazawa:2018b}. We derived explicit formulas for these kernels by the method of skew-orthogonal polynomials. The analysis at large matrix dimension will be carried out in~\cite{Kanazawa:2018b} and a summary of these results has been reported in~\cite{Kanazawa:2018}, since the calculations are very technical. Inspired from physics we particularly study the hard edge in those two works and derive the corresponding non-linear $\sigma$-model which is the chiral perturbation theory in QCD.

The considered random matrix interpolates between GUE ($\mu=0$) and chGUE ($\mu=1$) as does the model for the Hermitian Wilson Dirac operator in~\cite{Akemann:2011,Damgaard:2010cz,Akemann:2010em}. However our model preserves chirality at any time while it is broken in~\cite{Akemann:2011,Damgaard:2010cz,Akemann:2010em}. This leads to a non-uniform convergence in the limit $\mu\to0$ about the origin while it is uniform for the Hermitian Wilson Dirac operator (when the quark mass is vanishing), cf.~\cite{Akemann:2011,Damgaard:2010cz,Akemann:2010em}.   {The exact chiral pairs of eigenvalues $(\lambda_j,-\lambda_j)$ are the reason}, which repel each other   {the strongest} at the origin. This repulsion is absent for the Hermitian Wilson Dirac operator~\cite{Akemann:2011,Damgaard:2010cz,Akemann:2010em}. The implications of this behavior to applications in QCD will be studied in more detail in~\cite{Kanazawa:2018b}.

When considering our results in the limit $\mu\to0$ one has to be careful with the interpretation. The limit does not exactly yield the eigenvalue statistics of GUE but its singular value statistics. Thus one has to compare the results rather with those in~\cite{Akemann:2001,Forrester:2006,Edelman:2014,Bornemann:2016}.  The difference of the statistics is the sign of the eigenvalues over which one has to average. This yields two independent eigenvalue spectra equivalent to those of two GAOE ({Gaussian distributed imaginary antisymmetric matrices,} see~\cite{Kieburg:2017rrk} for the notation), one of even and one of odd dimension. The extension of the present   {model} to orthogonal and symplectic ensembles   {might be} an interesting direction of future research.

\section*{Acknowledgements}

 {We thank Gernot Akemann for given us comments on the first draft. Moreover, we} acknowledge support by the RIKEN iTHES project (TK) and  the German research council (DFG) via the CRC 1283: ``Taming uncertainty and profiting from randomness and low regularity in analysis, stochastics and their applications" (MK).  

\appendix

\section{New Unitary Group Integral}\label{app:group}

We consider the following integral 
\begin{equation}
\eqalign{
	\mathcal{I} & \equiv 	\int\limits_{\U(N)} d \mu(U)\; 
	\exp \left[\xi \Tr(AU+U^\dagger B)
	+ \frac{1}{2}\Tr [(AU)^2 + (U^\dagger B)^2] \right],
	\label{eq:Idefin}
}
\end{equation}
where $\xi\in\mathbb{C}$ is an arbitrary constant, 
$A$ and $B$ are complex $N\times N$ matrices, and 
$ d  \mu$ is the normalized Haar measure on the unitary group $\U(N)$. This group integral is slightly more general than we need in the main text. For $|\xi|\to\infty$ with $\xi A$ and $\xi B$ fixed we obtain the Leutwyler-Smilga integral~\cite{Leutwyler:1992yt}.

The integral $\mathcal{I}$ is invariant under $(A,B)\to(AU_0,U_0^\dagger B)$ for all $U_0\in{\rm U}(N)$. Thus it effectively depends on the product  $AB$, only, which is a crucial first step for the ensuing computations. When this simplification would not have worked we would have been lost. Without loss of generality we assume that 
the eigenvalues of $AB$ are non-degenerate. The degenerate case can be obtained at the end by employing l'Hospital's rule. We can also assume that $A=B=a=\diag(a_1,\ldots,a_N)\in\mathbb{C}^N$ since the integral~\eref{eq:Idefin} only depends on invariants of $AB$ due to the invariance under $AB\to V_0ABV_0^\dagger$ for all $V_0\in{\rm U}(N)$. Thus the matrix $AB$ can be diagonalized by a similarity transformation  {(we exclude Jordan blocks of size bigger than one)}. Moreover $a$ can be chosen real for the calculation below because the integral is analytic in $a$.

By means of a Hubbard-Stratonovich transformation with a Hermitian matrix $H$ we have
\begin{equation}
\eqalign{
	\mathcal{I} & = e^{-\Tr a^2}
	\int_{\U(N)} d\mu(U)\exp\left[\xi \Tr(aU+U^\dagger a) + \frac{1}{2}\Tr(aU+U^\dagger a)^2\right]
	\\
	& = \frac{e^{-\Tr a^2}}{2^{N/2}\pi^{N^2/2}}\int_{\Herm(N)} d H\exp\left[-\frac{1}{2} \Tr(H-\xi\1_{N})^2\right]\\
	& \quad \times\int_{\U(N)}  d  \mu(U)\exp\left[\Tr H(aU+U^\dagger a)\right].
}
\end{equation}
In the next step we perform the spectral decomposition $H=V^\dagger xV$ 
for $V\in\U(N)$ and $x=\diag(x_1,\dots,x_N)\in\mathbb{R}^N$ without  ordering. The measure 
becomes \cite{Mehta_book,Edelman_Rao_2005}
\begin{equation}
\eqalign{
	d H = \frac{\pi^{N(N-1)/2}}{N!\prod^{N-1}_{j=0}j!}  d \mu(V) 
	\Delta_N^2(x) \prod_{i=1}^{N}d x_i.
}
\end{equation}
Then the group integral~\eref{eq:Idefin} is
\begin{equation}\label{a.1}
\eqalign{
	\mathcal{I} =& \frac{e^{-\Tr a^2}}{(2\pi)^{N/2}N!\prod^{N-1}_{j=0}j!}
	\int_{\mathbb{R}^N} dx \Delta_N^2(x) \exp\left[-\frac{1}{2}\Tr(x-\xi\1_{N})^2\right]\\
	&\times
	\int_{[\U(N)]^2} d\mu(V)d\mu(U)\;
	\exp\left[\Tr V^\dagger xV(aU+U^\dagger a)\right].
}
\end{equation}
Shifting $U\to UV$ we notice that the group integral is the Berezin-Karpelevich integral~\cite{BK,Wettig},
\begin{equation}\label{BK-int}
\fl
\int_{[\U(N)]^2} d\mu(V)d\mu(U)
	 {\exp\left[\Tr (xVaU+U^\dagger aV^\dagger x)\right]}=\prod_{j=0}^{N-1}[j!]^2\ \frac{\det[I_0(2a_{j}x_k)]_{j,k=1,\ldots,N}}{\Delta_{N}(a^2)\Delta_{N}(x^2)}.
\end{equation}
We plug this integral into Eq.~\eref{a.1} and find
\begin{equation}\label{a.2}
\eqalign{
	\fl \mathcal{I} & = \frac{\prod^{N-1}_{j=0}j!}{(2\pi)^{N/2}N!}\frac{e^{-\Tr a^2}}{\Delta_N(a^2)}
	\int_{\mathbb{R}^N} dx \frac{\Delta_N^2(x)}{\Delta_N(x^2)}\exp\left[-\frac{\Tr(x-\xi\1_{N})^2}{2}\right]\det[I_0(2a_lx_k)]_{k,l=1\ldots,N}.
}
\end{equation}
Now we make use of the Pfaffian identity~\eref{Pfaff-ident} and after de Bruijn's integration theorem~\cite{deBruijn} we finally arrive at the result
\begin{equation}
\eqalign{
	\label{eq:evenI}
	\mathcal{I} & = \left(\prod_{j=0}^{N-1}\frac{j!}{\sqrt{4\pi}}\right)
	\frac{e^{-\Tr a^2}}{\Delta_N(a^2)}
	{\Pf}\left[\bm{B}_{\xi}(a_k,a_l)\right]_{k,l=1,\ldots, N}
}
\end{equation}
for $N$ even and
\begin{equation}
\eqalign{
	\label{eq:oddI}
	\mathcal{I} & = 
	\left(\prod_{j=0}^{N-1}\frac{j!}{\sqrt{4\pi}}\right)
		\frac{e^{-\Tr a^2}}{\Delta_N(a^2)}
	\Pf \left[\begin{array}{c|c} \bm{B}_{\xi}(a_k,a_l) & \bm{C}_\xi(a_k) \\ \hline - \bm{C}_\xi(a_l) & 0 \end{array}\right]_{k,l=1,\ldots,N}
}
\end{equation}
for $N$ odd. The weights in the Pfaffians are
\begin{eqnarray}
\eqalign{
	\fl \bm{B}_{\xi}(a_l,a_k) &= 
	\int_{\mathbb{R}^2} d x\,d y~ 
	\frac{x-y}{x+y} 
	\Big[ I_0(2a_l x)I_0(2a_k y)-I_0(2a_k y)I_0(2a_l x) \Big]
	e^{-[(x-\xi)^2+(y-\xi)^2]/2},
	\label{eq:defiB}
	\\
	\fl \bm{C}_\xi(a_k) &= \sqrt{2}\int_{-\infty}^{\infty}d x\; 
	e^{- (x-\xi)^2/2} I_0(2a_k x).
\label{eq:defBCap}
}
\end{eqnarray}

As pointed out before, taking the limit $\xi\to\infty$ with $a\xi\to\widehat{a}$ fixed yields the Leutwyler-Smilga integral~\cite{Leutwyler:1992yt}. For $N=2$ and $\widehat{a}=\diag(m_1,m_2)$ it becomes
\begin{equation}
\eqalign{
		\mathcal{I} &= \frac{1}{4\pi}\frac{1}{m_2^2-m_1^2}\lim_{\xi\to\infty}\xi^2\bm{B}_{\xi}(m_1/\xi ,m_2/\xi )\\ 
		&=  \frac{1}{4\pi}\frac{1}{m_1^2 - m_2^2}\lim_{\xi\to \infty}\xi^4
		\int_{\mathbb{R}^2} dx dy~ \frac{x-y}{x+y+2}\exp\left[-\frac{1}{2}\xi^2(x^2+y^2)\right]\\
		&\quad \times\Big[ I_0(2m_1 (x+1))I_0(2m_2 (y+1)) - I_0(2m_1 (y+1))I_0(2m_2 (x+1)) \Big]\\
		&=  \frac{1}{m_1^2 - m_2^2}[ m_1 I_1(2m_1)I_0(2m_2) - m_2 I_0(2m_1)I_1(2m_2) ], 
}
\end{equation}
	which agrees with \cite{Brower:1981vt,Jackson:1996jb,Akuzawa1998}. In the second line we substituted $(x,y)\to \xi(x+1,y+1)$ and performed a saddle point approximation in the last line.

We are interested in the opposite limit when $\xi\to0$. Then we can simplify the two integrals to
\begin{eqnarray}
	\fl \bm{B}_0(a_l,a_k) &=
	4 e^{a_l^2+a_k^2}
		\int_0^{\pi} d \vartheta\, 
		\tan \vartheta \sinh\big[ (a_l^2-a_k^2)\sin (2\vartheta) \big] 
		\; I_0(2a_la_k\cos (2\vartheta)),\quad
	\label{eq:defiB.b}
	\\
	\fl \bm{C}_0(a_k) &=2 \sqrt{\pi} e^{a_k^2} I_0(a_k^2)
\label{eq:defBCap.b}
\end{eqnarray}
with the help of Eqs.~(10.43.24) and~(10.43.28) in \cite{NIST:DLMF}.
For the case of odd matrix dimension $N$ we need the integral of these weights over one of the singular values. For the one point weight it is simply
\begin{equation}
\bar{g}=\int_0^\infty \exp\left[-\frac{\lambda^2}{2\mu^2}\right]\bm{C}_0(\sqrt{\eta_-}\lambda)\lambda d\lambda=2\sqrt{\pi}|\mu|.
\end{equation}
The integrated two point weight is slightly more involved and is
\begin{equation}
\eqalign{
\fl H(\lambda)&=\int_0^\infty d\lambda'\; \lambda' \lambda\exp\left[-\frac{\lambda'^2+\lambda^2}{2\mu^2}\right]\bm{B}_0(\sqrt{\eta_-}\lambda',\sqrt{\eta_-}\lambda)\\
\fl &=\lambda e^{-\eta_+\lambda^2}
		\int_0^{\pi} d \vartheta\tan \vartheta \biggl[\frac{1}{\eta_+-\eta_-\sin(2\vartheta)}\exp\left[\frac{\eta_--\eta_+\sin(2\vartheta)}{\eta_+-\eta_-\sin(2\vartheta)}\eta_-\lambda^2\right]\\
\fl		& \quad -\frac{1}{\eta_++\eta_-\sin(2\vartheta)}\exp\left[\frac{\eta_-+\eta_+\sin(2\vartheta)}{\eta_++\eta_-\sin(2\vartheta)}\eta_-\lambda^2\right]\biggl],
}
\end{equation}
where we employed  Eq.~(10.43.23) in \cite{NIST:DLMF}.

\section{Brief Review of Pfaffian Structures with a Mixture of Bi-Orthogonal and Skew-Orthogonal Polynomials}\label{app:Pfaff}

Let us emphasize that the notation chosen here is independent of the one from the main text though it is related to it. Moreover we describe the situation for spectra on the positive real line. However almost everything in this appendix carries over to discussions on the  {whole} real line and some of it even to spectra on the complex plane.

We want to consider a joint probability density on $\mathbb{R}_+$ of the general form
\begin{equation}\label{jpdf.gen}
p(\lambda)=\frac{C_N}{N!}\Delta_N(\lambda^2)\Pf\!\!\left[\begin{array}{c|c} G(\lambda_a,\lambda_b) & g_{c-1}(\lambda_a) \\ \hline  -g_{d-1}(\lambda_b) & 0 \end{array}\right]\underset{c,d=1,\ldots,n}{\underset{a,b=1,\ldots,N}{\ }}
\end{equation}
with $N+n$ even, $g_a$ some one-point weights, and  $G(\lambda_a,\lambda_b)=-G(\lambda_b,\lambda_a)$ being an anti-symmetric two-point weight. In the model of the main text we have only $n=0,1$.
The results which are derived here for these two cases are an alternative way to the method presented in~\cite[Chapter~5.5.]{Mehta_book}. The advantage of the present approach is that it is also true for $n>1$ where the method in~\cite[Chapter~5.5]{Mehta_book} fails.

Without loss of generality, the two-point weight shall be orthogonal to all polynomials of order $m-1$ in $\lambda^2$ and the one-point weight $g_{c-1}(\lambda_a)$ shall be orthogonal to all polynomials of order $c-2$. Here, the notion ``orthogonal" is equivalent to the following equation and, thence, its definition,
\begin{equation}
\langle \lambda^{2j}|f\rangle=0\quad{\rm and}\quad r_{c-1,l}=0
\end{equation}
for any integrable function $f$, $j=0,\ldots,m-1$ and $l=0,\ldots,c-2$ with $c=1,\ldots, n$, where we have defined the skew-symmetric product
\begin{equation}\label{skew-prod}
\langle f_1|f_2\rangle=\int_{\mathbb{R}_+^2}d\lambda_1 d\lambda_2\; G(\lambda_1,\lambda_2)f_1(\lambda_1)f_2(\lambda_2)
\end{equation}
and the moments
\begin{equation}
r_{c,l}=\int_0^\infty d\lambda\; g_c(\lambda) \lambda^{2l}.
\end{equation}
Then the normalization constant is
\begin{equation}
\eqalign{
C_N^{-1}&=\Pf\left[\begin{array}{c|c} \langle \lambda^{2(a-1)}|\lambda^{2(b-1)}\rangle & r_{c-1,a-1} \\ \hline  -r_{d-1,b-1} & 0 \end{array}\right]\underset{c,d=1,\ldots,n}{\underset{a,b=1,\ldots,N}{\ }}\\
&=(-1)^{n(n-1)/2}\prod_{j=0}^{n-1}\bar{g}_{j}\prod_{j=0}^{(N-n)/2-1}h_{2j+n}\;,
}
\end{equation}
where we have used
\begin{equation}
\bar{g}_{j}=r_{j,j}=(-1)^{j}\frac{C_{j}}{C_{j+1}},\ j=0,\ldots,n-1, \quad {\rm and}\quad h_{2j+n}=\frac{C_{2j+n}}{C_{2j+n+2}}.
\end{equation}
Then we construct monic polynomials $\{p_a\}_{a=0,\ldots,n-1}$ up to order $n-1$ which are bi-orthogonal to the weights $\{g_a\}_{a=0,\ldots,n-1}$ and pairs of skew-orthogonal monic polynomials $\{q_{2a+n},\widetilde{q}_{2a+n}\}_{a=0,\ldots,(N-n)/2-1}$ which are skew-orthogonal to the skew-symmetric product~\eref{skew-prod} and orthogonal to all one-point weights $g_a$. The explicit construction is~\cite{Kieburg:2012mix}
\begin{equation}\label{pol-construct}
\eqalign{
\fl p_j(\lambda^2)&=(-1)^{(j-1)(j-2)/2}C_{j}\det\left[\begin{array}{c|c} r_{b-1,a-1} & \lambda^{a-1} \end{array}\right]\underset{b=1,\ldots,j}{\underset{a=1,\ldots,j+1}{\ }},\\
\fl q_{2j+n}(\lambda^2)&=(-1)^nC_{2j+n}\Pf\left[\begin{array}{c|c|c} \langle \lambda^{2(a-1)}|\lambda^{2(b-1)}\rangle & r_{c-1,a-1} & \lambda^{2(a-1)} \\ \hline  -r_{d-1,b-1} & 0 & 0 \\ \hline -\lambda^{2(b-1)} & 0 & 0 \end{array}\right]\underset{ c,d=1,\ldots,n}{\underset{a,b=1,\ldots,2j+n+1 }{\ }},\\
\fl \widetilde{q}_{2j+n}(\lambda^2)&=(-1)^nC_{2j+n}\Pf\left[\begin{array}{c|c|c} \langle \lambda^{2(a-1)}|\lambda^{2(b-1)}\rangle & r_{c-1,a-1} & \lambda^{2(a-1)} \\ \hline  -r_{d-1,b-1} & 0 & 0 \\ \hline -\lambda^{2(b-1)} & 0 & 0 \end{array}\right]\underset{c,d=1,\ldots,n}{\underset{a,b=1,\ldots,2j+n,2j+2}{\ }}.
}
\end{equation}
Due to the multi-linearity and the skew-symmetry of the Pfaffian as well as the determinant, the orthogonality relations follow as well as the normalizations
\begin{equation}
\int_0^\infty d\lambda \; g_a(\lambda)p_b(\lambda^2)=h_a\delta_{ab}\quad{\rm and}\quad\langle q_{2c+n}|\widetilde{q}_{2d+n}\rangle=h_{2c+n}\delta_{cd}
\end{equation}
for $a,b=0,\ldots,n-1$ and $c,d=0,\ldots,(N-n)/2-1$.
When denoting the average of an observable $f$ with respect to a joint probability density~\eref{jpdf.gen} of dimensions $N$ and $n$ as
\begin{equation}
\langle f\rangle_{N,n}=\int_{\mathbb{R}_+^N}d\lambda \;p(\lambda) f(\lambda),
\end{equation}
the de Bruijn~\cite{deBruijn,KieburgGuhr:2010b,Kieburg:2012mix} and the Andr\'eief~\cite{Andreief,KieburgGuhr:2010a} identity can be applied backwards. In combination with the relations of the Vandermonde determinant,
\begin{equation}
\fl \det[\lambda_{a}^{2(b-1)}]_{a,b=1\ldots, j}=\Delta_j(\lambda^2)\quad{\rm and}\quad\det[\lambda_{a}^{2(b-1)}]\underset{b=1,\ldots,j-1,j+1}{\underset{a=1\ldots, j}{\ }}=\Delta_j(\lambda^2)\Tr\lambda^2,
\end{equation}
one finds the Heine-like formulas~\cite{Mehta_book,Akemann:2010mt}
\begin{equation}\label{pol-Heine}
\eqalign{
p_j(x^2)&=\langle \det(x^2\1_j-\lambda^2)\rangle_{j,j}\;,\\
q_{2j+n}(x^2)&=\langle \det(x^2\1_{2j+n}-\lambda^2)\rangle_{2j+n,n}\;,\\
\widetilde{q}_{2j+n}(x^2)&=\langle (x^2+\Tr \lambda^2)\det(x^2\1_{2j+n}-\lambda^2)\rangle_{2j+n,n}\;.
}
\end{equation}

Next, let us consider the partition function
\begin{equation}
Z_{N}^{(k_{\rm b},k_{\rm f})}(\kappa)=\left\langle\frac{\prod_{j=1}^{k_{\rm f}}\det(\kappa_{{\rm f},j}^2\1_{N}-\lambda^2)}{\prod_{j=1}^{k_{\rm b}}\det(\kappa_{{\rm b},j}^2\1_{N}-\lambda^2)}\right\rangle_N
\end{equation}
with $k_{\rm b}\leq k_{\rm f}+N$. The ratio of characteristic polynomials can be combined with the Vandermonde determinant in the joint probability density~\eref{jpdf.gen} as follows~\cite{KieburgGuhr:2010a,KieburgGuhr:2010b}
\begin{equation}\label{Berezinian}
\eqalign{
\fl &\Delta_N(\lambda^2)\frac{\prod_{j=1}^{k_{\rm f}}\det(\kappa_{{\rm f},j}^2\1_{N}-\lambda^2)}{\prod_{j=1}^{k_{\rm b}}\det(\kappa_{{\rm b},j}^2\1_{N}-\lambda^2)}\\
\fl =\,&(-1)^{k_{\rm b}(k_{\rm b}-1)/2}\frac{\prod_{a=1}^{k_{\rm b}}\prod_{b=1}^{k_{\rm f}}(\kappa_{{\rm b},a}^2-\kappa_{{\rm f},b}^2)}{\Delta_{k_{\rm b}}(\kappa_{\rm b}^2)\Delta_{k_{\rm f}}(\kappa_{\rm f}^2)}\det\left[\begin{array}{c|c} \lambda_b^{2(a-1)} & \kappa_{{\rm f},c}^{2(a-1)} \\ \hline \displaystyle\frac{1}{\kappa_{{\rm b},d}^2-\lambda_b^2} & \displaystyle\overset{\ }{\frac{1}{\kappa_{{\rm b},d}^2-\kappa_{{\rm f},c}^2}} \end{array}\right]\underset{d=1,\ldots,k_{\rm b}}{\underset{c=1,\ldots,k_{\rm f}}{\underset{b=1,\ldots,N}{\underset{a=1,\ldots,N+N_{\rm f}}{\ }}}}
}
\end{equation}
with $N_{\rm f}=k_{\rm f}-k_{\rm b}$. Note, we have a different sign convention for the Vandermonde determinant in~\cite{KieburgGuhr:2010a,KieburgGuhr:2010b}; here it is $\Delta_j(x)=\prod_{a<b}(x_b-x_a)$.  A generalized version of the de Bruijn identity~\cite{deBruijn,KieburgGuhr:2010b,Kieburg:2012mix} yields a $(N+n+2k_{\rm f})$-dimensional Pfaffian
\begin{equation}
\eqalign{
\fl &Z_{N}^{(k_{\rm b},k_{\rm f})}(\kappa)=(-1)^{k_{\rm b}(k_{\rm b}-1)/2+k_{\rm f}(k_{\rm f}-1)/2}C_N\frac{\prod_{a=1}^{k_{\rm b}}\prod_{b=1}^{k_{\rm f}}(\kappa_{{\rm b},a}^2-\kappa_{{\rm f},b}^2)}{\Delta_{k_{\rm b}}(\kappa_{\rm b}^2)\Delta_{k_{\rm f}}(\kappa_{\rm f}^2)}\\
\fl &\times\Pf\left[\begin{array}{c|c|c|c} \langle \lambda^{2(a-1)}|\lambda^{2(b-1)}\rangle & r_{c-1,a-1} & \langle \lambda^{2(a-1)}|\frac{1}{\kappa_{{\rm b},d}^2-\lambda^{2}}\rangle & \kappa_{{\rm f},e}^{2(a-1)} \\ \hline -r_{f-1,b-1} & 0 & \hat{g}_{f-1}(\kappa_{{\rm b},d}) & 0 \\ \hline \langle \frac{1}{\kappa_{{\rm b},g}^2-\lambda^{2}}|\lambda^{2(b-1)}\rangle & -\hat{g}_{c-1}(\kappa_{{\rm b},g}) &  \langle \frac{1}{\kappa_{{\rm b},g}^2-\lambda^{2}}|\frac{1}{\kappa_{{\rm b},d}^2-\lambda^{2}}\rangle & \frac{1}{\kappa_{{\rm b},g}^2-\kappa_{{\rm f},e}^2} \\ \hline -\kappa_{{\rm f},h}^{2(b-1)} & 0 & -\frac{1}{\kappa_{{\rm b},d}^2-\kappa_{{\rm f},h}^2} & 0\end{array}\right]\underset{e,h=1\ldots,k_{\rm f}}{\underset{d,g=1,\ldots,k_{\rm b}}{\underset{c,f=1,\ldots,n}{\underset{a,b=1,\ldots,N+N_{\rm f}}{\ }}}}
}
\end{equation}
with $\hat{g}_{c-1}(\kappa_{{\rm b},d})=\int_0^\infty d\lambda\; g_{c-1}(\lambda)/(\kappa_{{\rm b},d}^2-\lambda^2)$ the Cauchy transform of $g_{c-1}$. In the next step one uses the following identity
\begin{equation}\label{Pfaffian.identity}
\Pf\left[\begin{array}{cc} A & B \\ -B^T & C\end{array}\right]=\Pf[C+B^TA^{-1}B]\Pf[A]
\end{equation}
for two arbitrary even-dimensional antisymmetric matrices $A$ and $C$ and an arbitrary (rectangular)  matrix $B$. For simplicity we chose $N_{\rm f}=k_{\rm f}-k_{\rm b}$ to be even.  We have
\begin{equation}
\eqalign{
\fl&Z_N^{(k_{\rm b},k_{\rm f})}(\kappa)=\frac{C_N}{C_{N+N_{\rm f}}}\frac{\prod_{a=1}^{k_{\rm b}}\prod_{b=1}^{k_{\rm f}}(\kappa_{{\rm b},a}^2-\kappa_{{\rm f},b}^2)}{\Delta_{k_{\rm b}}(\kappa_{\rm b}^2)\Delta_{k_{\rm f}}(\kappa_{\rm f}^2)}\\
\fl&\times\Pf\left[\begin{array}{c|c} \frac{C_{N+N_{\rm f}}}{C_{N+N_{\rm f}+2}}(\kappa_{{\rm b},b}^2-\kappa_{{\rm b},a}^2)Z_{N+N_{\rm f}+2}^{(2,0)}(\kappa_{{\rm b},a},\kappa_{{\rm b},b}) & Z_{N+N_{\rm f}}^{(1,1)}(\kappa_{{\rm b},a},\kappa_{{\rm f},c})/(\kappa_{{\rm b},a}^2-\kappa_{{\rm f},c}^2) \\ \hline -Z_{N+N_{\rm f}}^{(1,1)}(\kappa_{{\rm b},b},\kappa_{{\rm f},d})/(\kappa_{{\rm b},b}^2-\kappa_{{\rm f},d}^2) & \frac{C_{N+N_{\rm f}}}{C_{N+N_{\rm f}-2}}(\kappa_{{\rm f},d}^2-\kappa_{{\rm f},c}^2)Z_{N+N_{\rm f}-2}^{(0,2)}(\kappa_{{\rm f},d},\kappa_{{\rm f},c}) \end{array}\right]
}
\end{equation}
with the indices $a,b=1,\ldots,k_{\rm b}$ and $c,d=1,\ldots,k_{\rm f}$. The kernels are obtained for particular choices of $k_{\rm b}$ and $k_{\rm f}$.
One can get another, more explicit representation of the three kernels when choosing the polynomials~\eref{pol-construct} instead of the monomials,
 \begin{eqnarray}
\fl &\frac{C_{N+N_{\rm f}}}{C_{N+N_{\rm f}+2}}(\kappa_{{\rm b},b}^2-\kappa_{{\rm b},a}^2)Z_{N+N_{\rm f}+2}^{(2,0)}(\kappa_{{\rm b},a},\kappa_{{\rm b},b})\nonumber\\
\fl =&\left\langle\left. \frac{1}{\kappa_{{\rm b},b}^2-\lambda^{2}}\right|\frac{1}{\kappa_{{\rm b},a}^2-\lambda^{2}}\right\rangle\nonumber\\
\fl &+\sum_{j=0}^{(N+N_{\rm f}-n)/2-1}\frac{1}{h_{2j+n}}\left(\left\langle q_{2j+n}\left|\frac{1}{\kappa_{{\rm b},a}^2-\lambda^{2}}\right.\right\rangle\left\langle \widetilde{q}_{2j+n}\left|\frac{1}{\kappa_{{\rm b},b}^2-\lambda^{2}}\right.\right\rangle-\{\kappa_{{\rm b},a}\leftrightarrow\kappa_{{\rm b},b}\}\right),\nonumber\\
\fl &\frac{Z_{N+N_{\rm f}}^{(1,1)}(\kappa_{{\rm b},a},\kappa_{{\rm f},c})}{\kappa_{{\rm b},a}^2-\kappa_{{\rm f},c}^2}\nonumber\\
\fl =&\frac{1}{\kappa_{{\rm b},a}^2-\kappa_{{\rm f},c}^2}+\sum_{j=0}^{n-1}\frac{p_j(\kappa_{{\rm f},c}^2)\widehat{g}_j(\kappa_{{\rm b},a})}{\bar{g}_j}+\sum_{j=0}^{(N+N_{\rm f}-n)/2-1}\nonumber\\
\fl &\times\frac{1}{h_{2j+n}}\left(q_{2j+n}(\kappa_{{\rm f},c}^2)\left\langle \widetilde{q}_{2j+n}\left|\frac{1}{\kappa_{{\rm b},a}^2-\lambda^{2}}\right.\right\rangle-\widetilde{q}_{2j+n}(\kappa_{{\rm f},c}^2)\left\langle q_{2j+n}\left|\frac{1}{\kappa_{{\rm b},a}^2-\lambda^{2}}\right.\right\rangle\right),\nonumber\\
\fl &\frac{C_{N+N_{\rm f}}}{C_{N+N_{\rm f}-2}}(\kappa_{{\rm f},d}^2-\kappa_{{\rm f},c}^2)Z_{N+N_{\rm f}-2}^{(0,2)}(\kappa_{{\rm f},d},\kappa_{{\rm f},c})\nonumber\\
\fl =&\sum_{j=0}^{(N+N_{\rm f}-n)/2-1}\frac{q_{2j+n}(\kappa_{{\rm f},c}^2)\widetilde{q}_{2j+n}(\kappa_{{\rm f},d}^2)-q_{2j+n}(\kappa_{{\rm f},d}^2)\widetilde{q}_{2j+n}(\kappa_{{\rm f},c}^2)}{h_{2j+n}}.
\end{eqnarray}
 Note that the two-point weight $G(\lambda_1,\lambda_2)$ is orthogonal to each polynomial of order $n-1$ which explains why some terms vanish.
 
As the last quantity, we wish to consider the $k$-point correlation function of the joint probability density~\eref{jpdf.gen},
\begin{equation}
R_{N}^{(k)}(\lambda_1,\ldots,\lambda_k)=\frac{N!}{(N-k)!}\int d\lambda_{k+1}\cdots d\lambda_N\;p(\lambda).
\end{equation}
Let us  {highlight} those variables over which we integrate by  {renaming them as}  $x_{j}$. Then, the $k$-point correlation function can be rewritten as
\begin{equation}
\eqalign{
\fl R_{N}^{(k)}(\lambda)&=\frac{C_N}{(N-k)!}\int_{\mathbb{R}_+^{N-k}}dx\;\Delta_N(\lambda^2,x^2)\\
\fl &\quad \times\Pf\left[\begin{array}{c|c|c} G(\lambda_a,\lambda_b) & G(\lambda_a,x_c) & g_{d-1}(\lambda_a) \\ \hline G(x_e,\lambda_b) & G(x_e,x_c) & g_{d-1}(x_e) \\ \hline  -g_{f-1}(\lambda_b) & -g_{f-1}(x_c) & 0 \end{array}\right]\underset{d,f=1,\ldots,n}{\underset{ c,e=k+1,\ldots,N }{\underset{a,b=1,\ldots,k}{\ }}}\\
\fl &=\frac{C_N}{(N-k)!} \sum_{s_1,\ldots, s_k=\pm1}\prod_{j=1}^k \frac{s_j}{2\pi i}\int_{\mathbb{R}_+^{N}}dx\; \Delta_N(\lambda^2,x_{k+1}^2,\ldots,,x_{N}^2)\\
\fl & \quad \times \prod_{j=1}^{k}\frac{1}{(\lambda_j-i s_j\epsilon)^2-x_j^2}\Pf\left[\begin{array}{c|c} G(x_a,x_b) & g_{c-1}(x_a) \\ \hline  -g_{d-1}(x_b) & 0 \end{array}\right]\underset{c,d=1,\ldots,n}{\underset{a,b=1,\ldots,N}{\ }}.
}
\end{equation}
The Pfaffian is antisymmetric in $x$ such that we can antisymmetrize also the other terms which yields a factor $(N-k)!/N!$ and a phase factor by combining the Vandermonde determinant with the Cauchy factors $1/((\lambda_j-i s_j\epsilon)^2-x_j^2)$,
\begin{equation}
\eqalign{
\fl &R_{N}^{(k)}(\lambda)=(-1)^{k}\frac{C_N}{N!} \lim_{\epsilon\to0}\sum_{s_1,\ldots, s_k=\pm1}\prod_{j=1}^k \frac{s_j}{2\pi i}\int_{\mathbb{R}_+^{N}}dx\\
\fl & \times \det\left[\begin{array}{c|c} x_a^{2(b-1)} & \displaystyle \frac{1}{(\lambda_c-i s_c\epsilon)^2-x_a^2} \\ \hline \lambda_d^{2(b-1)} & 0 \end{array}\right]\underset{c,d=1,\ldots,k}{\underset{a,b=1,\ldots,N}{\ }}\Pf\left[\begin{array}{c|c} G(x_a,x_b) & g_{c-1}(x_a) \\ \hline  -g_{d-1}(x_b) & 0 \end{array}\right]\underset{c,d=1,\ldots,n}{\underset{a,b=1,\ldots,N}{\ }}.
}
\end{equation}
We again use a generalized de Bruijn integral~\cite{deBruijn,KieburgGuhr:2010b,Kieburg:2012mix} and find after using the identity~\eref{Pfaffian.identity}
\begin{equation}
\eqalign{
R_{N}^{(k)}(\lambda)=&(-1)^{k(k-1)/2} \Pf\left[\begin{array}{c|c}
			W_{N}(\lambda_a,\lambda_b) & G_{N}(\lambda_a,\lambda_c)
			\\ \hline
			- G_{N}(\lambda_d,\lambda_b) & K_{N}(\lambda_d,\lambda_c) \end{array}\right]_{a,b,c,d=1,\ldots,k},
}
\end{equation}
where the kernels are
\begin{eqnarray}
\fl W_{N}(\lambda_a,\lambda_b)&=&\lim_{\epsilon\to0}\sum_{s_1,s_2=\pm1}\frac{s_1s_2}{(2\pi)^2}\frac{C_{N}}{C_{N+2}}((\lambda_a-i s_1\epsilon)^2-(\lambda_b-i s_2\epsilon)^2)\nonumber\\
\fl &&\times Z_{N+2}^{(2,0)}(\lambda_a-i s_1\epsilon,\lambda_b-i s_2\epsilon)\\
\fl &=&-G(\lambda_a,\lambda_b)+\sum_{j=0}^{(N-n)/2-1}\frac{1}{h_{2j+n}}\int_{\mathbb{R}_+^2}dx_1dx_2\;G(x_1,\lambda_a)G(x_2,\lambda_b)\nonumber\\
\fl &&\times\left(q_{2j+n}(x_1^2)\widetilde{q}_{2j+n}(x_2^2)-q_{2j+n}(x_2^2)\widetilde{q}_{2j+n}(x_1^2)\right),\nonumber\\
\fl G_{N}(\lambda_a,\lambda_c)&=&\lim_{\epsilon\to0}\sum_{s=\pm1}\frac{s}{2\pi i}\frac{Z_{N}^{(1,1)}(\lambda_a-i s\epsilon,\lambda_c)-1}{(\lambda_a-i s\epsilon)^2-\lambda_c^2}\nonumber\\
\fl &=&\sum_{j=0}^{n-1}\frac{p_j(\lambda_c^2)g_j(\lambda_a)}{\bar{g}_j}+\sum_{j=0}^{(N-n)/2-1}\frac{1}{h_{2j+n}}\int_0^\infty dx\; G(x,\lambda_a)\\
\fl &&\times( q_{2j+n}(\lambda_c^2)\widetilde{q}_{2j+n}(x^2)- q_{2j+n}(x^2)\widetilde{q}_{2j+n}(\lambda_c^2)),\nonumber\\
\fl K_{N}(\lambda_d,\lambda_c)&=&\frac{C_{N}}{C_{N-2}}(\lambda_d^2-\lambda_c^2)Z_{N+N_{\rm f}-2}^{(0,2)}(\lambda_d,\lambda_c)\nonumber\\
\fl &=&\sum_{j=0}^{(N-n)/2-1}\frac{q_{2j+n}(\lambda_c^2)\widetilde{q}_{2j+n}(\lambda_d^2)-q_{2j+n}(\lambda_d^2)\widetilde{q}_{2j+n}(\lambda_c^2)}{h_{2j+n}}.
\end{eqnarray}
The normalized level density of the joint probability density~\eref{jpdf.gen} takes then a rather simple form
\begin{equation}
\eqalign{
\fl&\rho_N(\lambda)=\frac{1}{N}R_{N}^{(1)}(\lambda)=\frac{1}{N}G_{N}(\lambda,\lambda)\\
\fl&=\sum_{j=0}^{n-1}\frac{p_j(\lambda^2)g_j(\lambda)}{N\bar{g}_j}+\int_0^\infty dx\;G(x,\lambda)\hspace*{-0.3cm}\sum_{j=0}^{(N-n)/2-1}\frac{q_{2j+n}(\lambda^2)\widetilde{q}_{2j+n}(x)- q_{2j+n}(x)\widetilde{q}_{2j+n}(\lambda^2)}{Nh_{2j+n}}.
}
\end{equation}
We want to emphasize that this representation is true for any $n,N\in\mathbb{N}$ with $n+N$ being even. It is a combination of general bi-orthogonal ensembles~\cite{Borodin,Mehta_book} and skew-orthogonal polynomials~\cite{Mehta_book}.

\end{document}